\documentclass[11pt]{article}

\usepackage[a4paper,margin=25mm]{geometry}
\usepackage[T1]{fontenc}
\usepackage{iftex}
\ifPDFTeX
  \usepackage[utf8]{inputenc}
\fi
\usepackage{lmodern}
\usepackage{amsmath,amssymb,bm}
\usepackage{graphicx}
\usepackage{booktabs,tabularx,array}
\usepackage{float}
\usepackage[numbers,sort&compress]{natbib}
\usepackage{xcolor}
\usepackage[colorlinks=true,allcolors=blue!55!black]{hyperref}
\usepackage{orcidlink}

\allowdisplaybreaks
\title{Hidden amplitude and dynamical information in phase-separation spectra}\author{%
\textit{Boliang Yu}\textsuperscript{1}\,\orcidlink{0009-0006-4664-5068},
\textit{Ruixin Zhou}\textsuperscript{1}\,\orcidlink{0009-0004-9303-6674}, and
\textit{Jiaxing Yuan}\textsuperscript{2}\thanks{Corresponding author: \texttt{jiaxingyuan@hkust-gz.edu.cn}}\,\,\orcidlink{0000-0001-9890-4961}\\[0.55em]
\parbox{0.90\textwidth}{%
\centering\small\itshape
\textsuperscript{1}School of Physics and Astronomy, Shanghai Jiao Tong University, Shanghai 200240, China\\
\textsuperscript{2}Advanced Materials Thrust, Function Hub,
The Hong Kong University of Science and Technology (Guangzhou),
Guangzhou 511453, China
}%
}

\date{}

\begin{document}
\maketitle

\begin{abstract}

Framewise variance normalization removes fluctuation amplitude from an observation, but not from the dynamics that generate later patterns. The current normalized spectrum therefore need not determine its own future. We study this effect in three two-dimensional phase-separation models that change transport, add active driving, or suppress large-scale separation. For paired initial fields, their normalized spectra evolve identically in the linear regime. The restoring model provides an exact example: damping does not directly change how the normalized spectrum evolves when both equations are applied to the same field. It nevertheless changes the hidden amplitude, which alters the later nonlinear shape. An exact comparison separates the immediate effect of changing the equation from the effect of reaching a different state through earlier evolution. It also reveals strong cancellation between these effects in the mobility and active models. As a practical test, spectral histories improve classification by about 4--6 percentage points at several later observation windows, although the gain depends on quench depth and spectral construction. Temporal spectra can therefore reveal dynamical consequences of information absent from every individual normalized frame.

\end{abstract}

\section{Introduction}\label{sec:introduction}

Phase separation begins as a transport problem. For a conserved scalar composition, gradients of chemical potential drive a flux, and the continuity equation converts that flux into an evolving spatial pattern \cite{CahnHilliard1958,HohenbergHalperin1977}. A quench into the spinodal region makes the homogeneous mixture unstable: the negative curvature of the homogeneous free energy amplifies long-wavelength fluctuations, while the square-gradient penalty suppresses sufficiently short wavelengths and selects a finite fastest-growing scale \cite{Cahn1961}. As the fluctuations grow, domains and interfaces form. Their later coarsening depends on both the thermodynamic driving force and the way composition is transported.

The equal-time structure factor, defined as the Fourier transform of a two-point correlation function, provides a compact description of this pattern evolution \cite{Bray1994,Furukawa1984}. In a regime controlled by one characteristic length, it has the scaling form $S_{\mathrm{conv}}(k,t)=L(t)^d f[kL(t)]$, where $d$ is the spatial dimension (here $d=2$) and $f$ is the scaled spectral shape. The peak tracks a characteristic domain size. However, similar scaled shapes do not imply identical transport. Simulations dominated by bulk diffusion and simulations dominated by surface diffusion can produce closely similar scaled structure factors even when their growth laws differ \cite{PuriBrayLebowitz1997}. Thus neither visual similarity nor a similar equal-time spectrum is sufficient, by itself, to identify a unique dynamical mechanism.

The time at which the pattern is observed adds a second difficulty. The windows examined here are not assumed to lie in an asymptotic scaling regime. Conserved phase separation can exhibit substantial preasymptotic corrections and crossovers caused by interfacial transport and by the ensemble of initial fluctuations \cite{Huse1986,CastellanoZannetti1996}. The same clock time can therefore correspond to different dynamical stages at different quench depths. We treat this finite-time dependence as part of the inference problem rather than as a failure of scaling theory.

Framewise normalization creates a sharper loss of state information. Rescaling each field to unit variance removes the root-mean-square amplitude of composition fluctuations, while Fourier power and radial averaging also discard phase and angular structure. We ask whether the resulting normalized spectrum is sufficient to determine its later normalized evolution. If a discarded variable still enters the nonlinear dynamics, two fields with the same normalized observation can subsequently evolve differently.

We study this question in a controlled comparison built around deterministic Model B, the standard diffusive dynamics of a conserved composition field when hydrodynamic flow is neglected. The three modifications alter composition transport, add active driving, or suppress long-wavelength growth. Each candidate and reference trajectory begins from the same initial field, and the models and observation are specified before fitting. The task is discrimination among three prescribed model families under a deliberately compressed observation, rather than unrestricted discovery of a governing equation. Sparse equation discovery and image-based phase-field inference address richer observation problems in which more of the evolving state is retained \cite{BruntonProctorKutz2016,RudyEtAl2017,ZhaoEtAl2020,KiyaniEtAl2022}.

Projection-operator theory gives a general reason why a reduced observable need not form a closed dynamical state: variables removed from observation can continue to affect its future \cite{Zwanzig1961,Mori1965,ReinboldGrigoriev2019}. Here this idea serves as motivation rather than as an inferred memory equation. Cahn--Hilliard--Oono dynamics provides an especially clear example: its restoring term disappears from the instantaneous normalized spectral rate at a fixed field, yet changes the amplitude that controls later nonlinear evolution. This suggests that a sequence of normalized spectra may contain distinctions absent from its final member.

We construct those three models around deterministic Model B. Model B is the standard passive description of diffusion-dominated spinodal decomposition and coarsening in binary mixtures or alloys when hydrodynamic flow is neglected \cite{CahnHilliard1958,HohenbergHalperin1977}. It provides a paired reference trajectory and is not a fourth classification label. The first modification makes the mobility depend on composition. This changes the local transport coefficient while preserving the passive free-energy landscape, a choice used when diffusivity varies with composition or when bulk and interfacial transport are compared \cite{PuriBrayLebowitz1997,BrayEmmott1995,CahnElliottNovickCohen1996}. The second modification is Active Model B (AMB), which adds a gradient term to the chemical potential that cannot be obtained as the derivative of a free energy. It is a minimal scalar model for broken detailed balance in phase-separating active matter \cite{StenhammarEtAl2013,WittkowskiEtAl2014}. The third is Cahn-Hilliard-Oono (CHO) dynamics. It adds a nonlocal contribution that becomes a linear restoring term in the evolution equation and suppresses long-wavelength growth. Its equilibrium free-energy construction follows Ohta and Kawasaki, whereas the corresponding restoring dynamics belongs to the Oono model lineage developed for frustrated microphase formation in block copolymers \cite{OhtaKawasaki1986,OonoShiwa1987,OonoBahiana1988,LiuGoldenfeld1989}.

These choices change three distinct ingredients of a common field theory: the mobility that transports composition, the thermodynamic form of the chemical potential, and the growth of long-wavelength fluctuations. All three retain a conserved scalar field and the same periodic geometry. They also reduce to the same Model-B equation when the mobility parameter $\alpha$, the Active-Model-B coefficient $\lambda$, or the CHO restoring coefficient $\beta$ is set to zero. We use the same baseline coefficients and paired initial fields throughout, observe only variance-normalized radial spectra, and omit these identical zero-parameter trajectories from the classification labels. The sampled parameter values define a controlled comparison; they are not calibrated to particular materials, and the first through fourth values within different model families are not assumed to have equal physical effects.

We first derive the common linear evolution of the normalized spectra and then use CHO to show exactly how discarded amplitude changes later nonlinear shape. A paired rate decomposition extends the analysis to Mobility and AMB by separating a direct change of equation from the effect of reaching different states. Finally, we compare endpoint and history classifiers and test the result against input dimension, spectral construction, quench depth, and parameter level.

\section{Methods}\label{sec:methods}

\subsection{Candidate dynamics and paired ensemble}\label{sec:candidates}

We consider a scalar composition field $\phi(\mathbf{r},t)$ on a two-dimensional periodic square domain $\Omega$, write $|\Omega|$ for its area, and write $\bar\phi$ for the spatial mean of $\phi$. The simulations use $\bar\phi=0$, and all quantities are dimensionless. Our common reference is deterministic Model B \cite{CahnHilliard1958,HohenbergHalperin1977} with unit constant mobility and free energy

\begin{equation}
\mathcal{F}_0[\phi]=\int_{\Omega}\left(\frac{r}{2}\phi^2+\frac{1}{4}\phi^4+\frac{\kappa}{2}|\nabla\phi|^2\right)\,d\mathbf{r}.
\end{equation}

The corresponding chemical potential is $\mu_0$. We denote the right-hand side of the Model-B evolution equation by $F_0[\phi]$:

\begin{equation}
\mu_0=r\phi+\phi^3-\kappa\nabla^2\phi,\qquad F_0[\phi]=\nabla^2\mu_0.
\end{equation}

To see how thermodynamics and transport enter separately, write the conservation law as $\partial_t\phi=-\nabla\cdot\mathbf{J}$ with flux $\mathbf{J}=-M(\phi)\nabla\mu_0$. For periodic boundaries and any positive mobility $M(\phi)$,

\begin{equation}
\frac{d\mathcal{F}_0}{dt}=-\int_{\Omega}M(\phi)|\nabla\mu_0|^2\,d\mathbf{r}\leq 0.
\end{equation}

The free energy therefore supplies the thermodynamic driving force, while the mobility controls the rate and spatial route by which composition moves down that energy landscape. For $r<0$, the local term $r\phi^2/2$ curves downward near $\phi=0$, so the uniform zero state lies in the spinodal region. This destabilizing curvature competes with the stabilizing square-gradient term proportional to $\kappa$ and produces the band of growing Fourier modes derived in Section~\ref{sec:linear-blind-limit} \cite{Cahn1961}.

Each candidate changes one part of this reference dynamics. The composition-dependent-mobility model uses $M_{\alpha}(\phi)=(1+\alpha\phi^2)^{-1}$. This mobility remains strictly positive, so it preserves the same $\mathcal{F}_0$ and its dissipative structure. Composition-dependent mobility can change coarsening kinetics \cite{PuriBrayLebowitz1997,BrayEmmott1995}. The rational form used here does not, however, approach zero in the bulk phases and therefore does not reach the degenerate, bulk-immobile limit that produces pure surface-diffusion asymptotics \cite{CahnElliottNovickCohen1996}. Those studies motivate sensitivity to the transport law; we do not import their growth laws into the present parameter range.

Active Model B (AMB) instead keeps the mobility constant and adds $\lambda|\nabla\phi|^2$ to the chemical potential. This is the lowest-order scalar gradient contribution that cannot be written as the functional derivative of a free energy. It therefore breaks detailed balance at gradient order while retaining conservation of the scalar field \cite{StenhammarEtAl2013,WittkowskiEtAl2014}. We study only the deterministic, constant-mobility sector and do not interpret the sampled $\lambda$ values as a quantitative calibration to a particular active-particle suspension.

The third model is an Oono-type extension of the Cahn-Hilliard equation, abbreviated CHO. It conserves the spatial mean but adds the nonlocal chemical-potential contribution $\beta(-\nabla^2)^{-1}(\phi-\bar\phi)$. After the Model-B gradient flow is taken, this contribution appears in the evolution equation as the linear term $-\beta(\phi-\bar\phi)$. The equilibrium nonlocal free energy is related to the Ohta-Kawasaki construction, and the restoring dynamics follows the Oono model lineage \cite{OhtaKawasaki1986,OonoShiwa1987,OonoBahiana1988,LiuGoldenfeld1989}. By penalizing long wavelengths, the term frustrates continued coarsening toward system-scale domains. Here it defines a controlled candidate rather than a calibrated kinetic model of a specific block-copolymer melt. Appendix~\ref{app:fourier-cho} gives the nonlocal energy explicitly. Table~\ref{tab:candidates} lists the three evolution equations and the sampled nonzero values of their additional parameters.

\begin{table}[H]
\centering
\small
\begin{tabularx}{\textwidth}{>{\raggedright\arraybackslash}p{0.15\textwidth} >{\raggedright\arraybackslash}X >{\raggedright\arraybackslash}p{0.24\textwidth}}
\toprule
\textbf{Model} & \textbf{Evolution equation} & \textbf{Positive parameter values} \\
\midrule
Mobility & $\partial_t\phi=\nabla\cdot[(1+\alpha\phi^2)^{-1}\nabla\mu_0]$ & $\alpha=0.05,0.1,0.2,0.8$ \\
AMB & $\partial_t\phi=\nabla^2[\mu_0+\lambda\lvert\nabla\phi\rvert^2]$ & $\lambda=0.05,0.1,0.2,0.8$ \\
CHO & $\partial_t\phi=F_0[\phi]-\beta(\phi-\bar\phi)$ & $\beta=0.0025,0.005,0.01,0.08$ \\
\bottomrule
\end{tabularx}
\caption{Candidate dynamics. The common parameters are $\kappa=1$ and $r=-0.8$ or $-1.2$. We call $r=-0.8$ the shallower quench and $r=-1.2$ the deeper quench. The four parameter values are ordered only within each family; positions in these ordered lists do not imply equal physical perturbations across families.}
\label{tab:candidates}
\end{table}

When $\alpha$, $\lambda$, or $\beta$ is zero, the corresponding equation reduces to the same Model-B reference. Giving different family labels to these identical trajectories would create incompatible labels for the same example. We therefore exclude all zero-parameter trajectories from classification. Model B is retained only as a paired physical reference. For each quench depth and initial-condition seed, every candidate and its reference start from the same mean-centered Gaussian field, drawn with a nominal standard deviation of $0.02$. The classification dataset thus contains $3\times4\times2\times30=720$ candidate trajectories. These trajectories solve known equations; the task is to distinguish their three labels within this finite dataset, not to discover an unrestricted equation from data.

\subsection{Numerical trajectories and targeted validation}\label{sec:numerics}

The simulations use a $64\times64$ periodic grid with spacing $\Delta x=1$, time step $\Delta t=0.02$, and 1,600 steps. Frames are saved every 50 steps, including $t=0$, giving 33 observations through $t=32$. Randomness enters only through the paired initial conditions; the evolution contains no stochastic forcing or hydrodynamics. Time integration uses double precision, while saved fields are stored in single precision. Before we compare a candidate with its paired Model-B trajectory as described in Section~\ref{sec:diagnostics}, the saved fields are promoted to double precision. The ensemble contains 30 independent initial-condition seeds and 60 Model-B reference trajectories across the two quench depths.

We compute spatial derivatives in Fourier space and evaluate nonlinear products on the $64\times64$ simulation grid. Let $\mu_{\mathrm{loc}}^{\,n}=r\phi^n+(\phi^n)^3$ denote the local part of the chemical potential evaluated at step $n$. A hat denotes the unnormalized forward discrete Fourier transform, and for a wave vector $\mathbf{k}$ we write $k=|\mathbf{k}|$. The Model-B step treats the stabilizing fourth-order term implicitly and $\mu_{\mathrm{loc}}^{\,n}$ explicitly:

\begin{equation}
\widehat{\phi}_{\mathbf{k}}^{\,n+1}
=\frac{\widehat{\phi}_{\mathbf{k}}^{\,n}
-\Delta t\,k^2\widehat{\mu}_{\mathrm{loc},\mathbf{k}}^{\,n}}
{1+\Delta t\,\kappa k^4}.
\end{equation}

For $\mathbf{k}=\mathbf{0}$, the factor $k^2$ vanishes, so the zero Fourier coefficient is invariant under this update and the spatial mean is conserved. The reported runs start from a mean-centered field, with any explicit mean projection removing only round-off drift. For AMB, the active gradient term is added to the explicitly evaluated chemical potential. For the composition-dependent-mobility model, we first compute the constant-mobility Model-B right-hand side and then add the explicit correction $\nabla\cdot[(M(\phi)-1)\nabla\mu_0]$. The same construction is used when the right-hand side is evaluated on a saved field in Section~\ref{sec:diagnostics}. On a Fourier grid with an even number of points, successive real first derivatives and a direct spectral Laplacian can treat the Nyquist modes, the highest resolved spatial frequencies, differently. We therefore compute the common Model-B part in the same way as in the time step. This makes the zero-parameter mobility right-hand side identical to Model B on the complete grid.

For CHO, each time step first applies the Model-B update above, producing the provisional field $\widetilde\phi^{\,n+1}$, and then applies the exact decay of its fluctuation about the mean:

\begin{equation}
\phi^{n+1}=\bar\phi+e^{-\beta\Delta t}(\widetilde\phi^{\,n+1}-\bar\phi).
\end{equation}

If the provisional field is held fixed, this decay changes its amplitude but leaves the normalized spectral observation defined in Section~\ref{sec:observable} unchanged. Exact treatment of this substep does not make the complete nonlinear update exact: the combined method remains first order. All 240 CHO trajectories with positive $\beta$ in the classification dataset use this update. The same update is used for one-step comparisons in which the CHO and Model-B calculations start from the same saved field; we do not replace it by a different semi-implicit scheme with the restoring term in the denominator.

The targeted numerical checks use four initial conditions, both quench depths, the weakest and strongest parameter values, and times through $T=16$. They compare CHO time steps of 0.02, 0.01, and 0.005. For the other families they also test Fourier padding, in which nonlinear products are evaluated on a larger grid before the retained modes are truncated back to the $64\times64$ grid. This reduces the folding of unresolved high-frequency products into the retained modes, an error called aliasing. We also compare paired trajectories computed with different time steps. These checks test the stated algebra and bound numerical discrepancies in selected regimes. They are not a convergence study of the complete classification experiment with 30 initial-condition groups. Appendix~\ref{app:checks} reports the ratios, time restrictions, and limitations.

\subsection{Normalized spectral observation}\label{sec:observable}

Our observation applies the usual structure-factor calculation to a field normalized to unit variance in each frame. The conventional equal-time structure factor used in scaling analyses retains the intensity of the order-parameter fluctuations \cite{Bray1994,Furukawa1984}. Dividing each saved field by its own root-mean-square fluctuation amplitude removes that scale before its Fourier power is formed. Fourier power contains no phase information, and radial binning removes directional information. Let $N=64^2$ be the number of grid points. For a field with nonzero variance, define the mean-centered fluctuation $\delta\phi$, its root-mean-square amplitude $a$, and the unit-variance field $\eta$ by

\begin{equation}
\begin{aligned}
\delta\phi_{\mathbf{r}}&=\phi_{\mathbf{r}}-\bar\phi,
&a^2&=\frac{1}{N}\sum_{\mathbf{r}}\delta\phi_{\mathbf{r}}^2,\\
\eta_{\mathbf{r}}&=\frac{\delta\phi_{\mathbf{r}}}{a}.
\end{aligned}
\end{equation}

Here $\bar\phi$ is the spatial mean. With the unnormalized forward discrete Fourier transform used in the simulations, $|\widehat{\eta}_{\mathbf{k}}|^2/N$ is the usual single-frame estimate of the wavevector-resolved equal-time structure factor of $\eta$. To express the same spectrum in the normalized-power notation used below, let $P_{\mathbf{k}}=|\widehat{\delta\phi}_{\mathbf{k}}|^2$, $E=\sum_{\mathbf{q}\ne\mathbf{0}}P_{\mathbf{q}}$, and $p_{\mathbf{k}}=P_{\mathbf{k}}/E$ for $\mathbf{k}\ne\mathbf{0}$, with $p_{\mathbf{0}}=0$. Parseval's identity gives $E=N^2a^2$, and therefore $|\widehat{\eta}_{\mathbf{k}}|^2/N=Np_{\mathbf{k}}$. If the sets $B_j$ partition the complete square Fourier grid into 24 equally spaced radial bins, the radial average of this structure factor gives the measured coordinates

\begin{equation}
\begin{aligned}
S_j(t)&\equiv\frac{1}{|B_j|}\sum_{\mathbf{k}\in B_j}
\frac{|\widehat{\eta}_{\mathbf{k}}(t)|^2}{N}\\
&=\frac{N}{|B_j|}\sum_{\mathbf{k}\in B_j}p_{\mathbf{k}}(t),\\
x_j(t)&=\log[1+S_j(t)],\qquad j=1,\ldots,24.
\end{aligned}
\end{equation}

The first expression is the radial-bin average of the single-frame estimate of the equal-time structure factor of $\eta$; the second equality shows that it is exactly the original normalized-power calculation of $S_j$. The count $|B_j|$ includes the zero grid point when it lies in a bin, although mean centering sets its power to zero. This convention fixes the normalization of the first bin. Applying the same radial-bin calculation to the mean-centered original field $\delta\phi$ gives $a^2(t)S_j(t)$ in each frame. Under the same discrete Fourier and radial-bin conventions, the conventional ensemble-averaged radial structure factor is therefore $S_{\mathrm{conv},j}(t)=\langle a^2(t)S_j(t)\rangle_{\mathrm{ens}}$. This relation states precisely what the framewise variance normalization removes. Thus the 24-component vector $\mathbf{x}(t)=(x_1(t),\ldots,x_{24}(t))$ contains neither amplitude, Fourier phase, nor angular variation within a radial bin. The logarithm is an invertible change of coordinates for the nonnegative binned spectrum and therefore discards no additional information.

We compare two inputs constructed from the saved vectors at times $t_n=n\Delta t_{\mathrm{save}}$ through a final observation time $T$. We call the final 24-component vector $\mathbf{x}(T)$ the endpoint input. We call the ordered concatenation of every saved vector from $t=0$ through $T$ the history input. Their dimensions are 24 and $24n_T$, respectively, where $n_T=T/\Delta t_{\mathrm{save}}+1$. Here $\Delta t_{\mathrm{save}}=1$, and the history includes the initial frame. The word endpoint refers only to the last frame in the chosen observation window; it does not imply equilibrium or a stationary morphology.

The normalization safeguards were inactive for every analyzed frame, so the observation retained the scale invariance defined above. Their numerical thresholds and the observed extrema are reported in Appendix~\ref{app:checks}.

Figure~\ref{fig:morphology-observation} follows the information removed at each step. The real-space panels show the raw order parameter $\phi$, which retains amplitude, Fourier phase, and angular structure. The complete measurement procedure from $\phi$ to $\mathbf{x}$ first subtracts the mean and divides by the field's standard deviation, then removes phase and angular structure through radial averaging of Fourier power. The classifier receives neither the raw field nor the normalized image. It receives only $\mathbf{x}(t)$. The endpoint supplies $\mathbf{x}(T)$, while the history supplies the ordered sequence from $t=0$ through $T$. Earlier frames therefore provide more measurements in the same compressed coordinates; they do not restore the amplitude, phase, or angular information removed from each frame. The visibly weaker late-time contrast in the illustrated CHO row is expected because the restoring term damps the root-mean-square amplitude of the composition fluctuations. The shared raw-field color scale exposes this amplitude difference, which is removed before the normalized spectrum is formed. Any apparent difference in late-time domain arrangement is illustrative in this single trajectory; Figure~\ref{fig:normalized-spectral-overlap} tests normalized spectral-shape differences across all paired initial conditions.

\begin{figure}[H]

\centering

\includegraphics[width=\textwidth]{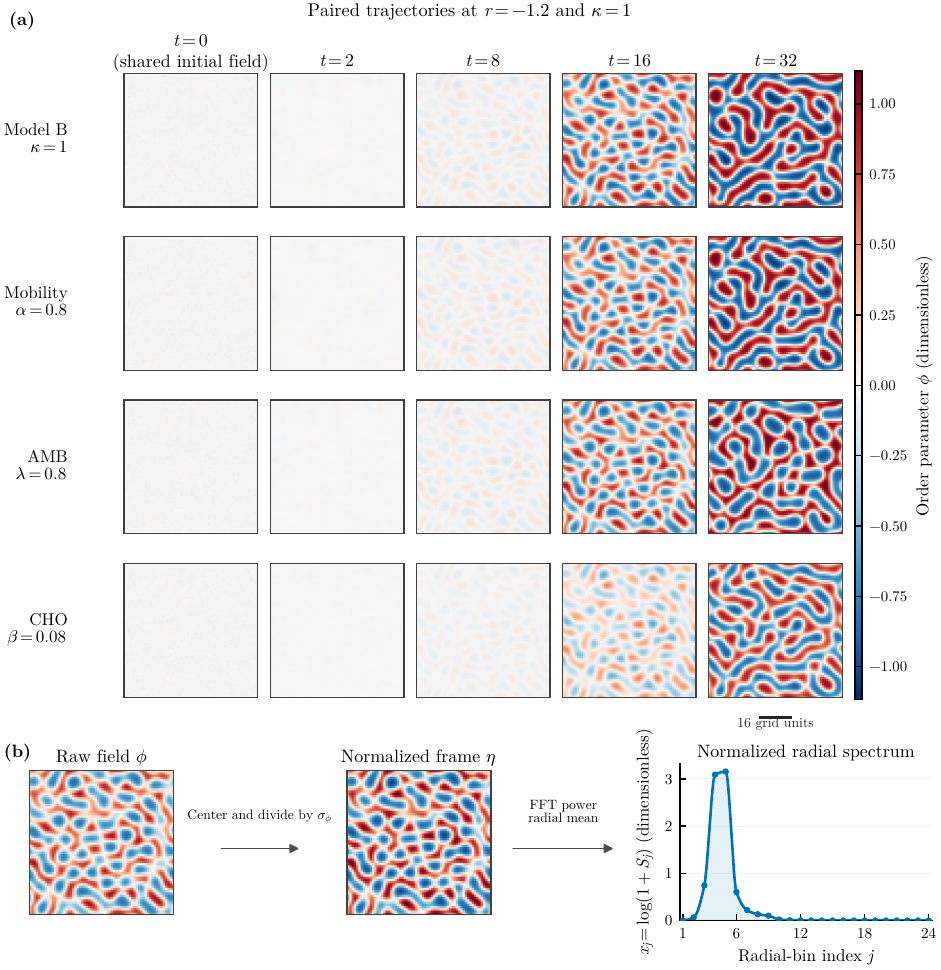}

\caption{\textbf{Paired morphologies and the observation map.} (a) Raw fields $\phi$ from one paired initial condition at $r=-1.2$ and $\kappa=1$. Rows show Model B and the strongest sampled Mobility ($\alpha=0.8$), AMB ($\lambda=0.8$), and CHO ($\beta=0.08$) trajectories; columns show $t=0,2,8,16,$ and 32. All panels share one color scale, so the weaker late-time CHO contrast shows its smaller fluctuation amplitude. The scale bar represents 16 grid units. (b) The measurement mean-centers each field, divides by its standard deviation to obtain $\eta$, forms the single-frame estimate $|\widehat{\eta}_{\mathbf{k}}|^2/N$ of its equal-time structure factor, and averages it into 24 radial bins before applying $x_j=\log(1+S_j)$. Symbols are the original bin values and the piecewise cubic curve is a visual guide only. The final 24-vector is the endpoint input; concatenating all saved vectors gives the history input. The raw fields are not supplied to the classifier.}

\label{fig:morphology-observation}

\end{figure}

\subsection{Spectral similarity and two sources of spectral change}\label{sec:diagnostics}

We use two complementary comparisons in the measured coordinates. The first asks how similar the binned spectra are at finite times. The second asks why two trajectories can have a small total rate difference even when changing the evolution equation at a fixed field has a larger effect. Both comparisons use each candidate and its Model-B reference generated from the same initial field and quench depth.

For the spectral-similarity comparison, we use the strongest sampled parameter in each family and all 30 paired initial conditions at each quench. At $T=16$, we first average each measured coordinate $x_j$ across those realizations. We also plot the signed difference in that same logarithmic coordinate, $\Delta x_j^{m}=\langle x_j^{m}-x_j^{\mathrm{B}}\rangle$, where the brackets here denote the mean over paired initial conditions and Model B is the reference. We then quantify realization-level differences before the logarithm. Define $w_j=|B_j|S_j/N$, the fraction of nonzero-mode power in bin $j$, so that $\sum_j w_j=1$. For a candidate and its paired Model-B reference, the Hellinger distance is $d_{\mathrm{H}}=[1-\sum_j\sqrt{w_j^{\mathrm{cand}}w_j^{\mathrm{ref}}}]^{1/2}$. It satisfies $0\leq d_{\mathrm{H}}\leq1$, with $d_{\mathrm{H}}=0$ for identical binned power distributions. We compute $d_{\mathrm{H}}$ at every saved time from 0 through 32 for each pair and then take the median and empirical 5th--95th percentiles across initial conditions. We impose no fitted threshold or equivalence test. Because the comparison uses the original bins and saved times, it retains differences in peak location and dynamical stage rather than aligning them away.

To examine instantaneous change, write the evolution equation for model $m$ as $\partial_t\phi=F_m[\phi]$. At a specified field $\phi$, let $v_m=F_m[\phi]$ denote the resulting rate field. The time derivative of the power in Fourier mode $\mathbf{k}$ is

\begin{equation}
R_{m,\mathbf{k}}[\phi]
=2\operatorname{Re}\left(\widehat{\phi}_{\mathbf{k}}^*
\widehat{v}_{m,\mathbf{k}}\right).
\end{equation}

In the zero-mean sector, differentiating the normalized modal powers gives

\begin{equation}
\dot p_{m,\mathbf{k}}[\phi]=\frac{R_{m,\mathbf{k}}-p_{\mathbf{k}}\sum_{\mathbf{q}\ne\mathbf{0}}R_{m,\mathbf{q}}}{E}.
\end{equation}

The subtraction removes the part of the power change that merely rescales the total nonzero-mode power. Unlike a logarithmic growth rate for one Fourier mode, this expression never divides by an individual $P_{\mathbf{k}}$. It therefore remains defined when a mode has zero power, provided that the total $E$ is positive. Differentiating the 24 measured coordinates then gives

\begin{equation}
G_{m,j}[\phi]=\frac{N}{|B_j|(1+S_j)}\sum_{\mathbf{k}\in B_j}\dot p_{m,\mathbf{k}}[\phi].
\end{equation}

The vector $G_m[\phi]$ is the instantaneous rate of change of the measured 24-component vector $\mathbf{x}$ when equation $m$ is evaluated on field $\phi$. This definition lets us distinguish a direct change of equation from the accumulated effect of arriving at a different field. Let $\phi_m(t)$ be a candidate field and $\phi_0(t)$ its paired Model-B field. At any saved time, both $F_m$ and $F_0$ can be evaluated on both $\phi_m$ and $\phi_0$. We symmetrically average these four evaluations. The \textbf{direct equation contribution}, denoted $\mathcal{O}_m$, changes the right-hand side while holding each field fixed:

\begin{equation}
\mathcal{O}_m=\frac{1}{2}\left(G_m[\phi_m]-G_0[\phi_m]+G_m[\phi_0]-G_0[\phi_0]\right),
\end{equation}

The \textbf{evolved-state contribution}, denoted $\mathcal{S}_m$, changes the field while holding each right-hand side fixed:

\begin{equation}
\mathcal{S}_m=\frac{1}{2}\left(G_m[\phi_m]-G_m[\phi_0]+G_0[\phi_m]-G_0[\phi_0]\right).
\end{equation}

Their sum is exactly the total difference between the rates of the two observed trajectories:

\begin{equation}
\mathcal{O}_m+\mathcal{S}_m=G_m[\phi_m]-G_0[\phi_0].
\end{equation}

This symmetric identity averages the two possible orders of changing the equation and changing the field. It is a specified accounting convention rather than a unique causal decomposition. Thus $\mathcal{S}_m$ measures the consequence of evaluating the equations on differently evolved fields within this convention; it is not a memory kernel. Appendix~\ref{app:checks} compares the two possible orders separately.

The two contributions can be individually large but oppositely directed. We describe their cancellation for each paired realization by

\begin{equation}
\mathcal{C}=1-\frac{\|\mathcal{O}_m+\mathcal{S}_m\|_2}{\|\mathcal{O}_m\|_2+\|\mathcal{S}_m\|_2}.
\end{equation}

The cancellation index $\mathcal{C}$ is defined whenever the denominator is nonzero and is calculated before averaging over initial conditions. It is zero when the two contributions reinforce one another and approaches one when two nonzero contributions nearly cancel. Because the index depends on both direction and relative norm, we also report the cosine $\rho_{\mathrm{op},state}$ between $\mathcal{O}_m$ and $\mathcal{S}_m$; values of $-1$, 0, and 1 denote opposite, orthogonal, and aligned vectors. Appendix~\ref{app:checks} defines an orthogonal reference and reports the sensitivity to the order of changing equation and field.

Every same-field and paired-field evaluation uses the same Fourier support, bin counts, and normalization. Each $G_m$ is obtained by applying the implemented spatial right-hand side to one saved field rather than by differencing sparsely saved frames. The inactive numerical floor and algebraic residuals are reported in Appendix~\ref{app:checks}.

For contribution magnitudes, we first compute the root-mean-square over the 24 coordinates for each realization and then average across initial conditions. Cancellation and directional quantities are also formed before averaging. The signed maps instead retain the sign of each coordinate when averaging over the 30 paired initial-condition groups, and their wave-number axis is scaled by $k_*=\sqrt{|r|/(2\kappa)}$. Bootstrap intervals resample these paired groups and describe variation across initial conditions, not numerical-discretization error. Appendix~\ref{app:checks} reports implementation residuals, order sensitivity, and the targeted discretization checks.

\subsection{Classification and grouped uncertainty}\label{sec:learning}

The physical comparisons above show how similar spectra and compensating instantaneous effects can arise. We next test whether earlier measurements in the same compressed coordinates improve discrimination among the three candidate labels. We fit multinomial logistic classifiers to either the endpoint input, the final 24-component vector, or the history input, the complete ordered sequence. All trajectories generated from the same initial-condition seed remain together in the training, validation, or test partition. This rule prevents trajectories evolved from the same initial field under different equations from appearing on opposite sides of a split. Five grouped folds provide one held-out prediction per trajectory for each input type and final observation time. Endpoint and history classifiers use the same classifier family, regularization grid, and group partitions; their input representations, and therefore their dimensions, differ.

For each final time $T=2,4,8,16,32$, five fixed folds divide the 30 initial-condition groups into 18 training, six validation, and six test groups. Every family, parameter value, and quench associated with a seed remains in the partition assigned to that seed, so related trajectories never appear in different partitions \cite{RobertsEtAl2017}. Input coordinates are standardized using the training data only. Multinomial logistic regression uses L2 regularization to limit coefficient magnitude and is optimized numerically with L-BFGS for at most 5,000 iterations. The inverse regularization strength is selected from $C_{\mathrm{reg}}=0.01,0.1,1,10$ by validation accuracy, with ties resolved by the order of this grid. After selection, the validation partition is not merged into the training partition. Neither the parameter value nor the quench depth is supplied as a classifier input.

We combine the five held-out folds before calculating statistics over initial-condition groups. Each group contributes 24 candidate trajectories when both quenches are pooled, or 12 when one quench is evaluated. Within each group, we calculate the history accuracy minus the endpoint accuracy and then average this paired difference over groups. Let $D(T)$ denote this mean accuracy difference at final time $T$. Percentile intervals are obtained from 10,000 bootstrap resamples of the complete initial-condition groups \cite{Efron1979}, which preserves the pairing between input types, model parameters, and quenches. The quench-specific intervals in the main text use the same 30 groups rather than fold-level aggregates. For the ten pairwise comparisons among $D(2)$, $D(4)$, $D(8)$, $D(16)$, and $D(32)$, we additionally apply the Holm step-down adjustment to control the familywise error rate across those comparisons \cite{Holm1979}. The other displayed intervals cover one reported quantity at a time and are not adjusted to form simultaneous confidence bands.

One control gives the history and endpoint inputs the same dimension. We first standardize the history using the training data, then use principal-component analysis to find 24 orthogonal directions of greatest training-set variance by full singular-value decomposition, without rescaling the resulting components to equal variance. The resulting 24 component scores are standardized again before logistic regression because their scales affect the L2 penalty. Both standardizers and the principal-component basis are fitted using training data only.

Further controls change the Fourier support and radial-bin boundaries, evaluate fixed predictions on parameter subsets, or refit after one parameter level within each family has been withheld from both training and validation. The within-family levels do not represent equal physical effects across models. Appendix~\ref{app:checks} describes an additional training-only selection of saved time--radial-bin coordinates using the magnitudes of $\mathcal O_m$ and $\mathcal S_m$. Each modified input is standardized on its own training data and its regularization is selected on validation data; test groups do not enter these choices.

\section{Results}\label{sec:results}

\subsection{Linear indistinguishability after frame normalization and finite-time spectral overlap}\label{sec:linear-blind-limit}

We first ask what the measured spectrum can distinguish while the field remains close enough to the uniform zero state for the equations to be linearized. For a nonzero Fourier mode, the Model-B amplitude grows exponentially at the rate

\begin{equation}
\sigma_0(k)=|r|k^2-\kappa k^4.
\end{equation}

The leading correction to the field rate is cubic in the Mobility family and quadratic in Active Model B (AMB). Neither correction therefore changes the linear growth rate. Cahn-Hilliard-Oono (CHO) dynamics does change that rate, giving $\sigma_\beta(k)=\sigma_0(k)-\beta$. To see what remains after each frame is normalized by its own total power, consider the normalized modal powers $p_{\mathbf{k}}$. Write $\sigma(k)$ for the relevant family's radial growth-rate function and $q=|\mathbf{q}|$. For real growth rates, the normalized powers satisfy

\begin{equation}
\dot p_{\mathbf{k}}=2\left(\sigma(k)-\sum_{\mathbf{q}\ne\mathbf{0}}p_{\mathbf{q}}\sigma(q)\right)p_{\mathbf{k}}.
\end{equation}

The uniform CHO shift $-\beta$ appears in both terms and cancels. The three families therefore produce identical normalized spectra in the linearized problem when they begin with the same normalized distribution of modal power and use the same common coefficients. The paired initial fields in this study satisfy that condition. CHO still changes the absolute growth of every mode, but frame normalization removes that common change from the spectral shape. This equality of normalized spectra along paired trajectories applies only to the linearized equations; nonlinear trajectories need not retain equal normalized spectra even when their spatial means remain zero.

Although normalization hides the uniform rate shift from the linear spectral shape, CHO still changes which modes grow. Its continuous unstable band is defined by $\sigma_{\beta}(k)>0$. When $\beta<r^2/(4\kappa)$, the growing wave numbers satisfy

\begin{equation}
\frac{|r|-\sqrt{r^2-4\kappa\beta}}{2\kappa}<k^2<\frac{|r|+\sqrt{r^2-4\kappa\beta}}{2\kappa}.
\end{equation}

All sampled values lie in this regime. The largest $\beta$ is 0.08, whereas $r^2/(4\kappa)$ is 0.16 for the shallow quench, $r=-0.8$, and 0.36 for the deep quench, $r=-1.2$. The restoring term narrows the unstable band and lowers every absolute modal growth rate, while leaving the fastest-growing wave number unchanged. Thus equality of the linear normalized shapes does not imply equality of either the instability bands or the amplitude histories.

For Model B, the fastest-growing wave number satisfies $k_*^2=|r|/(2\kappa)$. The inverse of its maximum growth rate is $\tau_0=4\kappa/r^2$, which gives 6.25 and 2.78 for the two quenches. CHO has the same fastest-growing wave number, but its maximum rate is lower: $\sigma_{\beta,\max}=r^2/(4\kappa)-\beta$. Its corresponding inverse growth rate is therefore $\tau_{\beta}=[r^2/(4\kappa)-\beta]^{-1}$ within the unstable regime. At the largest sampled $\beta$, these times are 12.5 and 3.57. Because a finite periodic grid need not contain a mode exactly at $k_*$, these continuous values do not specify the discrete evolution completely. They nevertheless show why a fixed final observation time can include different amounts of linear amplification at different quench depths and values of $\beta$. This motivates separate comparisons of the two quenches; it does not imply that their nonlinear classification curves collapse after time rescaling.

\begin{figure}[tbp]

\centering

\includegraphics[width=\textwidth]{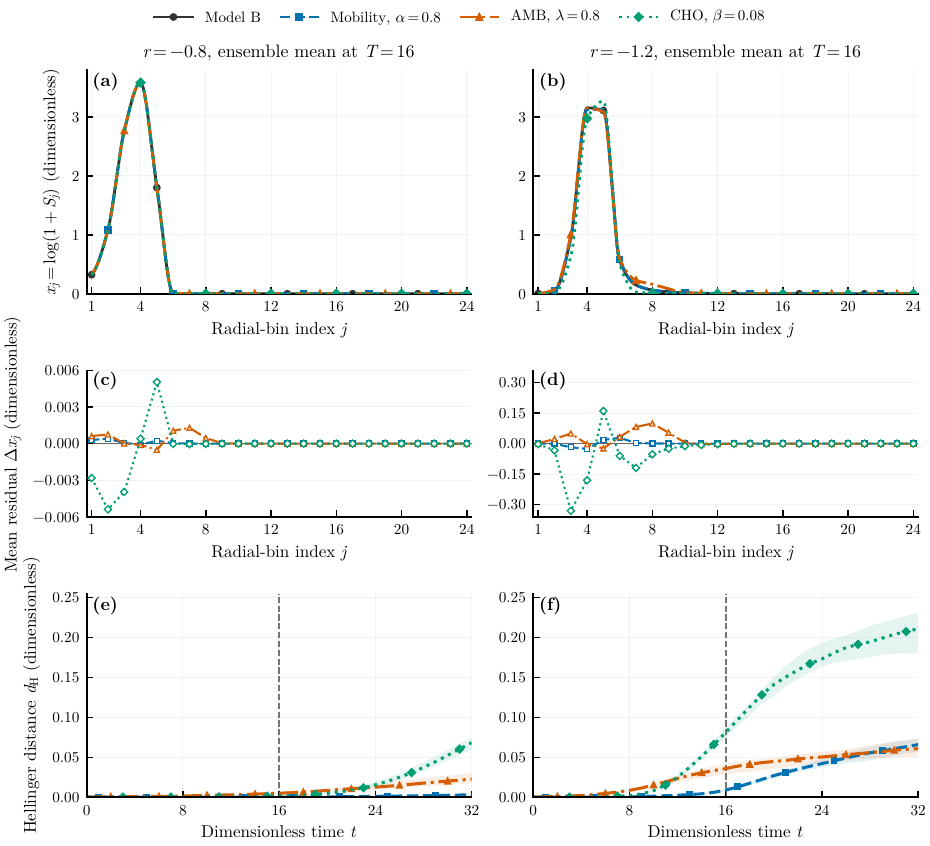}

\caption{\textbf{Normalized spectral shapes overlap strongly but not exactly.} (a,b) Ensemble-mean spectra at $T=16$ for Model B and the strongest sampled Mobility ($\alpha=0.8$), AMB ($\lambda=0.8$), and CHO ($\beta=0.08$) trajectories at the two quench depths. Symbols mark selected original bins; shape-preserving piecewise cubic curves guide the eye through all 24 bin means. (c,d) Candidate-minus-Model-B residuals in the measured logarithmic coordinates. Straight segments join the original bin values, and the panels use different labelled vertical scales. (e,f) Median paired Hellinger distances with empirical 5th--95th percentile bands over 30 initial-condition groups. Symbols mark saved times, straight segments connect adjacent values, and the dashed line marks $T=16$. All reported distances and classifier results use the original bin values and saved times rather than the display interpolation.}

\label{fig:normalized-spectral-overlap}

\end{figure}

The preceding equality is exact only while nonlinear corrections remain negligible. We therefore next compare the finite-time observations in the simulations. Figure~\ref{fig:normalized-spectral-overlap} shows the ensemble-mean measured coordinates at the highlighted final time $T=16$, their candidate-minus-Model-B residuals, and the paired Hellinger distances through time. Every curve uses the same 24 bins and the same clock time. No curve is shifted to align its peak, rescaled by an evolving length, or registered to a model-dependent time. The spectra for the shallow quench are nearly coincident. To expose differences that the absolute curves conceal, the middle row replots the same measured coordinates as signed residuals from the paired Model-B mean; it is not an additional classifier input. The two residual panels use separately labelled vertical scales because their magnitudes differ by about two orders of magnitude. In the deep quench, all spectra retain the same dominant band, but CHO shows a visible displacement.

To quantify the remaining differences for each paired initial condition, let $w_j=|B_j|S_j/N$ be the fraction of nonzero-mode power in bin $j$, so that $\sum_j w_j=1$. The Hellinger distance between a candidate and its paired Model-B reference is $d_{\mathrm{H}}=[1-\sum_j\sqrt{w_j^{\mathrm{cand}}w_j^{\mathrm{ref}}}]^{1/2}$. It lies between zero and one, with $d_{\mathrm{H}}=0$ for identical binned power distributions. At $T=16$, the largest shallow-quench median relative to Model B is 0.0050. For the deep quench, the corresponding medians are 0.0089, 0.0359, and 0.0812 for Mobility, AMB, and CHO; the central 90\% of the paired CHO values is [0.0748, 0.0869]. We also compare all four trajectories for each initial condition by taking the largest of their six pairwise distances. The median of this maximum is 0.0060 for the shallow quench and 0.1055 [0.0939, 0.1152] for the deep quench. The distances rise at later times, especially for deep-quench CHO. The finite-time spectra therefore overlap substantially, but they are neither equal nor related by a universal nonlinear collapse.

\subsection{How discarded amplitude affects later spectral shape}\label{sec:hidden-amplitude}

The linear result raises a second question: can amplitude affect the measured spectral shape even though the observation removes it from every saved frame? CHO gives a direct answer. Write the zero-mean field as $\phi=a\eta$, where $a$ is its root-mean-square amplitude and the normalized field $\eta$ satisfies $\langle\eta^2\rangle=1$; brackets denote a spatial average. Define $L=r\nabla^2-\kappa\nabla^4$ and $Q[\eta]=\nabla^2\eta^3$. The projection $\mathcal{P}_{\eta}v=v-\eta\langle\eta v\rangle$ removes the component of $v$ parallel to $\eta$, thereby preserving the unit-variance condition on $\eta$. Direct substitution into the CHO equation gives

\begin{equation}
\frac{\dot a}{a}=\langle\eta L\eta\rangle+a^2\langle\eta Q[\eta]\rangle-\beta,
\end{equation}

\begin{equation}
\dot\eta=\mathcal{P}_{\eta}\left(L\eta+a^2Q[\eta]\right).
\end{equation}

The restoring coefficient $\beta$ appears explicitly in the amplitude equation but not in the equation for the normalized field. The amplitude $a$, however, multiplies the nonlinear term that changes normalized shape. Two fields can therefore have the same $\eta$ but different values of $a$, and consequently different rates of normalized shape change whenever the projected cubic term is nonzero. Even the complete normalized field then fails to determine its own future rate without the discarded amplitude. The radial spectrum used here removes still more information by discarding Fourier phase and angular structure.

This mechanism does not require the damping term itself to change the normalized spectrum instantaneously when CHO and Model B are evaluated on the same field. Uniform damping changes only that field's amplitude. As trajectories evolve, however, damping changes the amplitude at which nonlinear redistribution takes place, so the later normalized shapes can separate. The transformation $\phi=e^{-\beta t}\psi$ expresses the same fact in another form: it moves the damping into a time-dependent coefficient multiplying $\nabla^2\psi^3$ (Appendix~\ref{app:fourier-cho}). Both expressions follow exactly from the continuous equation and were not fitted to the classification results. They show how discarded amplitude can influence the observation at later times, but they do not guarantee that a particular sequence of spectra or a particular classifier will recover that influence.

\subsection{Exact cancellation of the direct CHO effect at a fixed field}\label{sec:fixed-state-cancellation}

The amplitude argument predicts a stronger same-field result. Recall that $G_m[\phi]$ is the time derivative of the measured 24-component spectrum produced by evolution law $m$ when it is evaluated on field $\phi$. For CHO, the additional modal power rate is $R_{\beta,\mathbf{k}}-R_{0,\mathbf{k}}=-2\beta P_{\mathbf{k}}$. This rate changes each modal power in the same proportion, so its contribution is removed when the spectrum is divided by total power. Consequently,

\begin{equation}
G_{\beta}[\phi]=G_0[\phi]
\end{equation}

for every fixed field with nonzero variance, including nonlinear fields. This equality compares CHO and Model B on the same field. It does not compare the two laws along the different fields they produce, so it does not imply $G_{\beta}[\phi_{\beta}(t)]=G_0[\phi_0(t)]$. The first comparison isolates the immediate effect of changing the equation; the second includes the accumulated effect of earlier evolution. Distinguishing them is necessary before interpreting a direct contribution that vanishes. The numerical implementation satisfies this identity to roundoff accuracy, as documented in Appendix~\ref{app:checks}.

\subsection{Separating direct equation changes from the effects of evolved states}\label{sec:operator-state}

The symmetric decomposition defined in Methods makes the distinction from Section~\ref{sec:fixed-state-cancellation} quantitative along every paired trajectory. The direct equation contribution $\mathcal{O}_m$ measures the effect of changing the evolution law while holding the field fixed. The evolved-state contribution $\mathcal{S}_m$ measures the effect of applying fixed laws to fields separated by earlier evolution. Their exact sum is the difference between the two rates of change of the measured spectrum. The AMB addition $\lambda\nabla^2|\nabla\phi|^2$ is quadratic in the field gradients and therefore couples pairs of Fourier modes. When the spectrum is concentrated near $k_{\mathrm{peak}}$, this coupling can produce a broad response near $2k_{\mathrm{peak}}$; Appendix~\ref{app:mechanism-support} gives the Fourier-space derivation.

\begin{figure}[H]

\centering

\includegraphics[width=\textwidth]{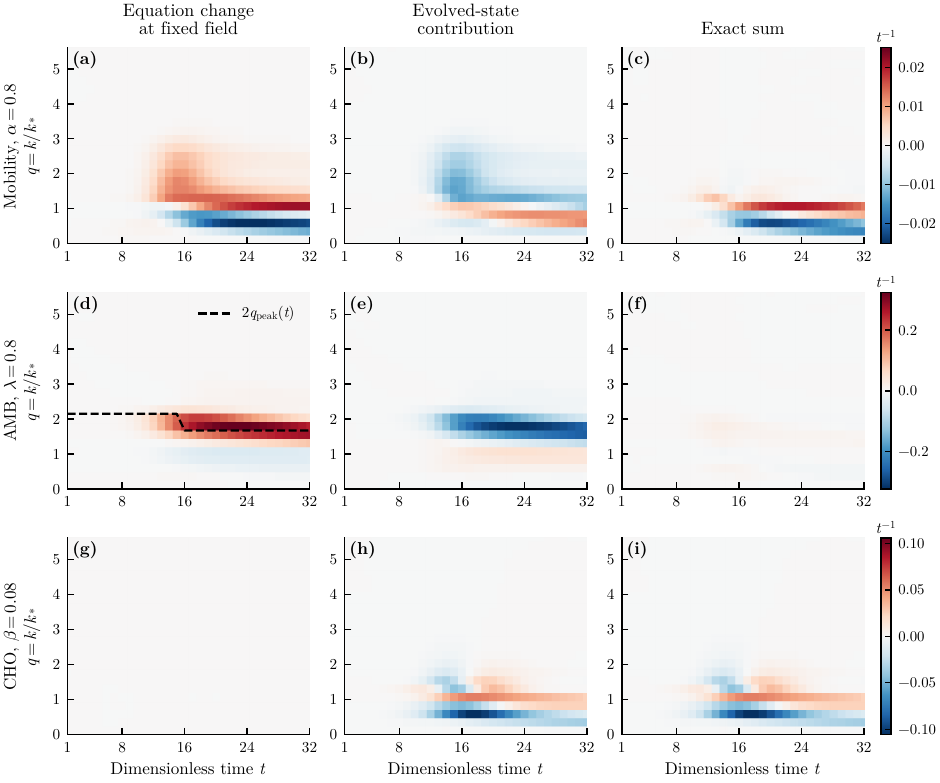}

\caption{\textbf{Direct equation and evolved-state contributions to normalized spectral evolution.} Rows show the strongest sampled Mobility ($\alpha=0.8$), AMB ($\lambda=0.8$), and CHO ($\beta=0.08$) perturbations at $r=-1.2$. Columns show the direct equation contribution $\mathcal{O}_m$, the evolved-state contribution $\mathcal{S}_m$, and their exact sum. Each map is the signed mean over 30 paired initial-condition groups at $t=1,\ldots,32$; wave number is scaled by $k_*=\sqrt{|r|/(2\kappa)}$, and each family row has one common color scale. The dashed AMB guide marks $2k_{\mathrm{peak}}$. The weak AMB sum results from opposition between two strong contributions, whereas the blank CHO $\mathcal{O}_m$ panel reflects exact cancellation of uniform modal decay under frame normalization.}

\label{fig:signed-budget}

\end{figure}

Figure~\ref{fig:signed-budget} shows that strong AMB produces direct-equation and evolved-state contributions with nearly equal magnitudes and opposite directions. At $t=16$, their mean root-mean-square magnitudes are 0.0707 and 0.0668, whereas their sum has magnitude 0.00592. The cancellation index is 0.956 and the directional cosine is $-0.998$, confirming that the small total results from opposition rather than from two individually weak terms. The broad direct response lies near the $2k_{\mathrm{peak}}$ scale expected from quadratic mode coupling, without implying an exact harmonic relation.

Mobility shows the same structure more weakly: at $t=16$, the two contributions have root-mean-square magnitudes 0.00664 and 0.00605, their sum is 0.00404, and their directional cosine is $-0.799$. CHO gives the limiting contrast. Its direct equation contribution vanishes exactly, so its evolved-state contribution equals the total rate difference. Detailed roundoff residuals, orthogonal references, and evaluation-order checks are reported in Appendix~\ref{app:checks}.

For continuous paired trajectories whose initial observations are identical, the difference between their final measured spectra is the time integral of the total rate difference $\mathcal{O}_m+\mathcal{S}_m$. Contributions can cancel between $\mathcal{O}_m$ and $\mathcal{S}_m$ at one time and can also cancel across different times. A small final separation is therefore compatible with distinct paths through the measured 24-dimensional space. The values evaluated at saved frames do not reconstruct the integral over every unsaved numerical step, and their norms do not determine a classifier's decision boundary. They instead provide a physical reason to test whether the sequence of saved spectra contains useful information beyond its final member.

\subsection{Does the sequence improve pooled classification?}\label{sec:endpoint-history}

\begin{figure}[tbp]

\centering

\includegraphics[width=\textwidth]{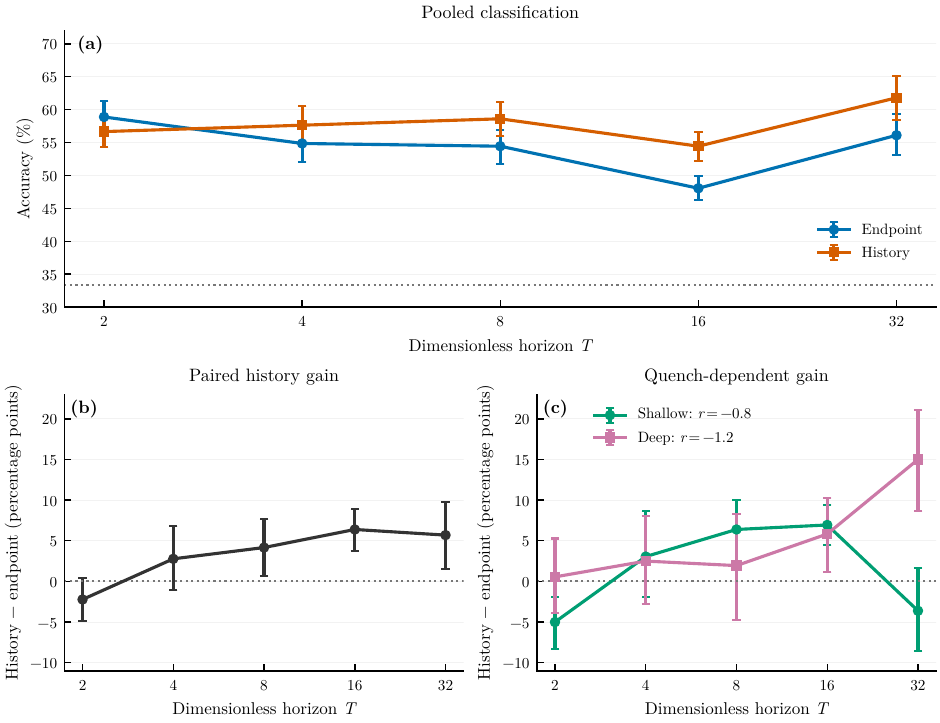}

\caption{\textbf{The benefit of earlier frames depends on time and quench depth.} (a) Pooled held-out classification accuracy; the dotted line marks one-third accuracy. (b) Paired history-minus-endpoint accuracy. (c) The same contrast at each quench depth. Panels (a,b) summarize 720 trajectories and panel (c) 360 per quench. Error bars are nominal paired 95\% intervals from 10,000 resamples of the 30 initial-condition groups, with preprocessing and trained classifiers held fixed. Symbols mark the sampled final times and straight segments guide the eye.}

\label{fig:history}

\end{figure}

We now make that statistical comparison. The endpoint input contains only the last saved 24-component spectrum, whereas the history input contains the ordered sequence of saved spectra from $t=0$ through the same final time $T$. Let $A_{\mathrm{history}}(T)$ and $A_{\mathrm{endpoint}}(T)$ be the held-out accuracies of the two classifiers, and define their paired difference as $D(T)=A_{\mathrm{history}}(T)-A_{\mathrm{endpoint}}(T)$. Figure~\ref{fig:history} shows that the sign and size of this gain depend on the observation window. At $T=2$, endpoint and history accuracies are 58.89\% and 56.67\%, respectively. At $T=16$, they are 48.06\% and 54.44\%, which gives $D(16)=6.39$ percentage points and a paired 95\% interval of [3.75, 8.89]. At $T=32$, the two accuracies are 56.11\% and 61.81\%, with a gain of 5.69 points [1.53, 9.72]. The nominal gain intervals at $T=8,16,32$ exclude zero; those at $T=2,4$ do not.

These intervals were obtained by resampling the 30 initial-condition groups with a percentile bootstrap \cite{Efron1979} while holding the preprocessing and trained classifiers fixed. They quantify variation across the grouped held-out predictions. They do not include variation that would arise from repeatedly choosing data splits, refitting preprocessing, or retuning the classifiers. Because a history also contains more input coordinates than an endpoint, the observed gain cannot yet be assigned solely to temporal information; Section~\ref{sec:controls} addresses that alternative.

The gain at $T=16$ is not resolved as larger than the gain at $T=32$. Specifically, $D(32)-D(16)=-0.69$ points, with interval [-5.56, 4.17]. The differences between $D(8)$, $D(16)$, and $D(32)$ and the earliest-window value $D(2)$ remain positive after applying the Holm adjustment to all ten pairwise comparisons between final observation times \cite{Holm1979}. Their adjusted bootstrap $p$ values are 0.0036, 0.0020, and 0.0048. Thus the evidence supports a change from the earliest window, but it does not identify one uniquely optimal final observation time.

\subsection{Why the result differs between quench depths}\label{sec:quench-window}

The pooled result combines regimes with different dynamics, so we next examine each quench separately. For the shallower quench at $T=16$, endpoint accuracy is exactly one third and history accuracy is 40.28\%, giving a gain of 6.94 points [4.44, 9.44]. This chance-level score describes the fitted endpoint classifier rather than equality of the underlying distributions. At the same final time, the deeper quench is more separable: endpoint and history accuracies are 62.78\% and 68.61\%, giving a gain of 5.83 points [1.11, 10.28].

The difference becomes sharper at $T=32$. The shallow-quench gain is -3.61 points [-8.61, 1.67], whereas the deep-quench gain is 15.00 points [8.61, 21.11]. A positive pooled value at this time therefore does not mean that waiting longer benefits both regimes. The mean endpoint score can recover while the earlier frames add a resolved positive gain for one quench but not for the other.

The quench-dependent linear amplification times in Section~\ref{sec:linear-blind-limit} provide one physical difference between the regimes. Section~\ref{sec:hidden-amplitude} then shows how CHO amplitude differences can alter subsequent normalized shape, while the fixed-scale fields in Figure~\ref{fig:morphology-observation} illustrate the associated contrast change for one paired realization. The decomposition in Section~\ref{sec:operator-state} also shows that, at the largest sampled parameter values, the deeper quench develops much larger differences in the time derivative of the measured spectrum by $t=16$. Together, these results make the quench dependence qualitatively plausible, but they do not quantitatively predict the accuracy curves. Classification also depends on the separation and orientation of the three feature distributions, the variation within each family, the observed coordinates, and the fitting procedure.

\subsection{Earlier spectra can retain information about the path}\label{sec:controls}

Sections 3.2--3.4 show why an instantaneous normalized spectrum need not provide a complete state description. In CHO, fields with the same normalized shape can have different amplitudes and therefore different subsequent nonlinear evolution. In Mobility and AMB, distinctions created along a trajectory can oppose the immediate effect of changing the equation. A final spectrum records only the remainder after these contributions have accumulated, whereas a sequence of spectra samples part of the path by which that endpoint was reached.

A history vector contains more coordinates than a single endpoint, so we tested whether the resolved gain at $T=16$ survives when the two inputs have equal dimension. Compressing each history to 24 principal components fitted on the training data gives a history-minus-endpoint gain of 4.17 percentage points [0.56, 7.78]. The gain at this window therefore cannot be explained only by the number of input coordinates. At $T=32$, however, the corresponding value is -0.69 points [-4.72, 3.33], so the same compression does not preserve an advantage at every observation time.

Earlier normalized spectra can therefore retain information about a trajectory that is weak in its final normalized spectrum, but the amount retained depends on when and how the system is observed. Changing the Fourier support changes the measured gain because it changes the projection of the evolving field. Appendix~\ref{app:checks} reports the full representation and held-parameter comparisons together with Figure~\ref{fig:controls}; parameter-subset results are given in Appendix~\ref{app:classifier-support}. Together, these comparisons show that the measured gain depends on the observation while leaving the physical distinction between an instantaneous normalized spectrum and the dynamics that determine its future unchanged.

\section{Discussion}\label{sec:discussion}

The observation map sends many physically different fields to the same normalized radial spectrum. Fields with the same normalized shape but different overall fluctuation amplitudes map to the same normalized field, while Fourier power and radial averaging further remove phase and angular structure. The central question is whether fields sharing this reduced description also share the same future spectral evolution. The CHO calculation shows explicitly that they can evolve differently. The normalized spectrum can therefore fail to provide a closed state description even while it records the instantaneous radial distribution of spectral power.

CHO reveals the mechanism particularly clearly. In the linear regime, the restoring term shifts the growth rate of every Fourier mode by the same amount. Frame normalization removes this common change, leaving the normalized spectral evolution identical to that of Model B for paired initial fields. The amplitude--shape decomposition shows what changes beyond this regime. Writing the field as $\phi=a\eta$ separates the root-mean-square amplitude $a$ from the unit-variance shape $\eta$. The restoring coefficient changes the evolution of $a$, while the nonlinear term governing $\eta$ is weighted by $a^2$. The amplitude trajectory can therefore separate before the normalized spectra visibly diverge. Once nonlinear mode coupling becomes appreciable, the altered amplitude changes the rate at which power is redistributed across wave numbers. Information removed from the observation has then re-entered the observed dynamics through the nonlinear evolution.

The direct-equation/evolved-state decomposition separates immediate equation effects from accumulated state effects in the Mobility and AMB families. The direct-equation contribution measures the immediate spectral effect of applying a different evolution law to the same field. The evolved-state contribution measures the effect of applying the equations to fields produced by their respective earlier trajectories. For the strong Mobility and AMB perturbations shown in Figure~\ref{fig:signed-budget}, the simulations give contributions with comparable magnitudes and opposite directions. Their sum can consequently be much smaller than either contribution. Since an endpoint difference accumulates the net spectral-rate difference over time, cancellation can occur both between the two contributions at a given time and between different stages of the trajectory. A spectral history samples the system before this accumulation is complete and can retain distinctions that have become weak in the final frame. For composition-dependent mobility, this offers a dynamical route by which different transport laws can generate closely similar two-point spectral shapes \cite{PuriBrayLebowitz1997}. For AMB, the evolved-state contribution can also carry differences in Fourier phase, angular structure, and interface geometry generated along the active trajectory.

The classifier comparison displays an observable consequence of this picture. At the pooled late windows $T=8,16,$ and 32, spectral histories raise accuracy by about 4--6 percentage points relative to the corresponding endpoints. The gain changes with quench depth and spectral construction, linking its visibility to both the dynamical stage and the observation used to view it. These comparisons place the result in the framework of practical identifiability: they measure distinguishability among the three prescribed model families through the normalized spectral observation \cite{WielandEtAl2021}.

The linear growth rates provide a physical basis for the dependence on quench depth. A deeper quench amplifies fluctuations more rapidly and brings the field into the nonlinear regime earlier, allowing hidden differences in amplitude and morphology to influence spectral redistribution sooner. A shallower quench spends a longer interval near the common linear evolution and reaches model-dependent nonlinear evolution later. The same clock time consequently represents different dynamical stages in the two quenches. The observed history gains are consistent with sequences that span different portions of this transition from common linear growth to model-dependent nonlinear evolution.

This interpretation also gives the result a direct experimental meaning. Experimental images are often normalized frame by frame to compensate for illumination, detector response, or sample-to-sample contrast. When image contrast is calibrated to the physical order parameter, that normalization removes a dynamical coordinate. Recording the frame variance alongside the normalized image retains this discarded image-scale amplitude at negligible storage cost. A time sequence may reveal dynamical consequences of variables absent from individual frames. Phase-sensitive, orientation-sensitive, or interface-based measurements could retain complementary information removed by the radial spectrum. The broader principle is that the usefulness of history is controlled by the relation between the observation map and the variables that drive nonlinear evolution.

The present comparison isolates the amplitude mechanism exactly in CHO and examines related evolved-state effects in Mobility and AMB within deterministic two-dimensional conserved-field dynamics. Among the three families, the quantitative AMB results depend most strongly on resolving the short-wavelength transfer generated by its active gradient term. Trajectories that resolve this nonlinear spectral transfer across the full wave-number range will sharpen the measured balance between direct-equation and evolved-state contributions. Restoring amplitude, phase, and directional information one coordinate at a time will reveal how the value of spectral history changes as the observed state becomes more complete. Extending the same analysis to noisy images, hydrodynamic phase separation, and experimentally measured trajectories would connect these mechanisms to richer forms of pattern-forming dynamics.

\section{Conclusion}\label{sec:conclusion}

Framewise variance normalization removes fluctuation amplitude from each observed spectrum while the underlying field continues to evolve with that amplitude. Starting from the observation map, the linear analysis showed that the three model families share the same normalized spectral evolution for paired initial fields in the linear regime. The CHO amplitude--shape decomposition then identified the missing dynamical coordinate. Its restoring term contributes exactly zero to the normalized spectral rate when CHO and Model B are evaluated on the same field, yet it changes the amplitude that controls subsequent nonlinear redistribution. The direct-equation/evolved-state decomposition separated immediate equation effects from the consequences of reaching different states, and the strong Mobility and AMB cases showed that these contributions can oppose one another.

For paired trajectories, an endpoint contains the time-integrated net spectral change, whereas a history also samples intermediate differences before some contributions cancel. At several pooled late windows, histories improve classification among the three model families, and the gain changes with quench depth and spectral construction. The principal physical conclusion is that information absent from every instantaneous normalized spectrum can remain active in the dynamics and become visible through time. Preserving a calibrated amplitude coordinate and retaining temporal observations therefore provide physically motivated routes for distinguishing phase-separation dynamics from compressed imaging data.

%

\section*{Data availability}

Data supporting this article are available on Zenodo: \url{https://doi.org/10.5281/zenodo.22307461}.

\section*{Acknowledgements}

The authors acknowledge the High-Performance Computing (HPC) facilities of The Hong Kong University of Science and Technology (Guangzhou) for providing computational resources and support.

\begingroup
\small
\bibliographystyle{unsrtnat}
\bibliography{references}
\endgroup
\clearpage

\appendix
\section{Fourier normalization and the CHO amplitude equation}\label{app:fourier-cho}

For the discrete transform used to construct the observation, the forward transform has no factor of $1/N$, whereas the inverse transform contains $1/N$. Parseval's identity for the mean-centered field therefore gives

\begin{equation}
\sum_{\mathbf{k}}|\widehat{\delta\phi}_{\mathbf{k}}|^2
=N\sum_{\mathbf{r}}|\delta\phi_{\mathbf{r}}|^2=N^2a^2.
\end{equation}

Because $\eta=\delta\phi/a$ and $p_{\mathbf{k}}=|\widehat{\delta\phi}_{\mathbf{k}}|^2/(N^2a^2)$ for nonzero modes, this identity gives $|\widehat{\eta}_{\mathbf{k}}|^2/N=Np_{\mathbf{k}}$. Its radial-bin average is exactly $S_j$ as defined in Section~\ref{sec:observable}.

The apparently local restoring term $-\beta\phi$ in the zero-mean CHO evolution comes from a nonlocal free-energy contribution. Ohta and Kawasaki introduced the corresponding equilibrium nonlocal free energy, whereas the Oono models introduced phenomenological Cahn--Hilliard dynamics with a linear restoring term; later work made the inverse-Laplacian representation explicit \cite{OhtaKawasaki1986,OonoShiwa1987,OonoBahiana1988,LiuGoldenfeld1989}. On a periodic domain, let $\mathcal{K}=(-\nabla^2)^{-1}$ denote the inverse Laplacian acting only on nonzero Fourier modes. With unit constant mobility, the free energy

\begin{equation}
\mathcal{F}_{\beta}=\mathcal{F}_0+\frac{\beta}{2}\int_{\Omega}\phi\mathcal{K}\phi\,d\mathbf{r}
\end{equation}

generates $\partial_t\phi=\nabla^2(\delta\mathcal{F}_{\beta}/\delta\phi)=F_0[\phi]-\beta\phi$. To display the nonlocal term in Fourier space, define the continuous Fourier coefficient by $\widehat{\phi}^{c}_{\mathbf{k}}=\int_{\Omega}\phi(\mathbf{r})e^{-i\mathbf{k}\cdot\mathbf{r}}d\mathbf{r}$. Then

\begin{equation}
\mathcal{F}_{\beta}-\mathcal{F}_0=\frac{\beta}{2|\Omega|}\sum_{\mathbf{k}\ne\mathbf{0}}\frac{|\widehat{\phi}^{c}_{\mathbf{k}}|^2}{k^2}.
\end{equation}

The area factor follows from this continuous Fourier convention; the superscript $c$ distinguishes these coefficients from the unnormalized discrete sums used in the measured features. The factor $1/k^2$ penalizes nonzero modes most strongly at long wavelength. Thus the restoring term has a clear thermodynamic effect even though, at one fixed field, it does not change the instantaneous rate of the variance-normalized spectrum.

The role of the discarded amplitude can also be seen directly in the CHO equation. For a zero-mean field, write $\phi=e^{-\beta t}\psi$ and define $L=r\nabla^2-\kappa\nabla^4$. The linear damping then cancels from the transformed equation, leaving

\begin{equation}
\partial_t\psi=L\psi+e^{-2\beta t}\nabla^2\psi^3.
\end{equation}

The normalized fields, and hence the normalized spectra, of $\psi$ and $\phi$ are identical at the same time. Nevertheless, the coefficient of the nonlinear term changes with time. This is the same amplitude dependence obtained from the normalized-field equation in Section~\ref{sec:hidden-amplitude}.

Finally, consider two continuous trajectories that start from the same field: $\phi_\beta(t)$ evolves under CHO and $\phi_0(t)$ under Model B. Let $H(\phi)=\mathbf{x}$ denote the ideal logarithmic-spectrum observation. In the next two equations, $x_\beta(t)=H[\phi_\beta(t)]$ and $x_0(t)=H[\phi_0(t)]$ each denote the complete 24-component observed vector, despite the unbold type used to keep the notation compact. For any perturbation $v$, $DH(\phi)[v]$ denotes the directional derivative of $H$ at $\phi$ in the direction $v$. Define the cubic part of the Model-B right-hand side by $N_3(\phi)=\nabla^2\phi^3$ and its contribution to the observed rate by $G_N(\phi)=DH(\phi)[N_3(\phi)]$. Positive scale invariance, $H(c\phi)=H(\phi)$, implies $DH(\phi)[\phi]=0$. Under $\phi\mapsto c\phi$, the contribution from the linear part of the right-hand side is unchanged, whereas $G_N$ scales as $c^2$. Therefore, if $G_0(\phi)=DH(\phi)[L\phi+N_3(\phi)]$ is the full Model-B rate of the observed vector, then $DG_0(\phi)[\phi]=2G_N(\phi)$. For sufficiently smooth trajectories from the same nonzero-variance field $\phi_{\mathrm{init}}$,

\begin{equation}
\left.\partial_t(x_{\beta}-x_0)\right|_{t=0}=0,\qquad \left.\partial_t^2(x_{\beta}-x_0)\right|_{t=0}=-2\beta G_N(\phi_{\mathrm{init}}),
\end{equation}

and hence

\begin{equation}
x_{\beta}(t)-x_0(t)=-\beta t^2G_N(\phi_{\mathrm{init}})+O(t^3).
\end{equation}

Writing $\phi_{\mathrm{init}}=a_{\mathrm{init}}\eta_{\mathrm{init}}$ gives $G_N(\phi_{\mathrm{init}})=a_{\mathrm{init}}^2G_N(\eta_{\mathrm{init}})$. The expansion applies directly to $\log(1+S)$ because the logarithm preserves the scale invariance of $H$. Its coefficient can vanish for special states, and its sign can differ among spectral coordinates. This is a local analytic result, not an empirically fitted scaling law or a guarantee that a classifier will benefit from earlier frames.

\section{Numerical and statistical checks}\label{app:checks}

\begin{figure}[H]

\centering

\includegraphics[width=\textwidth]{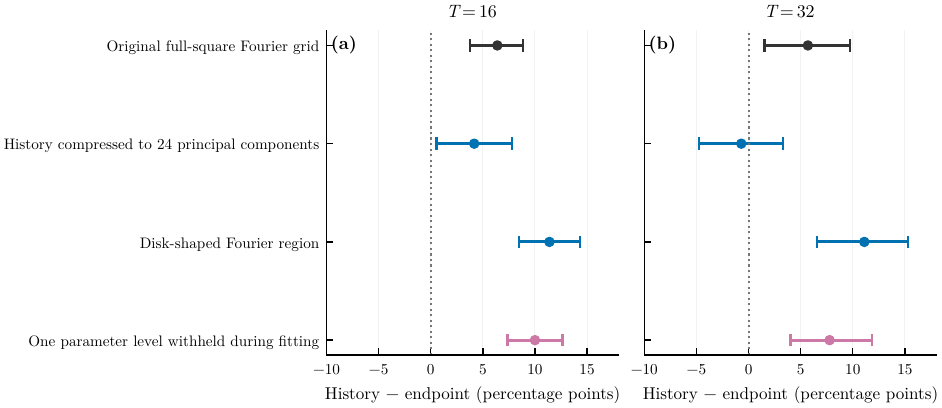}

\caption{\textbf{Dependence of the history comparison on representation and parameter sampling.} Points and horizontal bars show mean history-minus-endpoint accuracy and nominal 95\% initial-condition-group intervals at $T=16$ and 32. The rows compare the primary full-square representation with equal-dimensional principal-component compression, a disk-shaped Fourier region, and refitting after one parameter level within each family is withheld. The panels are supporting comparisons of distinct analysis choices.}

\label{fig:controls}

\end{figure}

\textbf{History representation and parameter checks.} Figure~\ref{fig:controls} collects the supporting comparisons removed from the main narrative. Compressing each history to 24 training-fitted principal components gives a history-minus-endpoint gain of 4.17 points [0.56, 7.78] at $T=16$ and -0.69 points [-4.72, 3.33] at $T=32$. This comparison gives the history and endpoint the same input dimension, although the compressed coordinates are not the same physical coordinates as the endpoint bins.

Restricting the Fourier support to $|\mathbf{k}|\leq\pi$ and rebinning that range gives an 11.39-point gain [8.47, 14.31] at $T=16$. The change affects both the retained modes and the bin boundaries, so it demonstrates sensitivity to the spectral construction rather than a separate causal role for the removed corner modes. The complete square grid remains the primary representation.

When one parameter level within each family is withheld from both training and validation, refitting on the remaining levels gives gains of 10.00 points [7.36, 12.64] at $T=16$ and 7.78 points [4.03, 11.81] at $T=32$. Within the sampled range, the history advantage is therefore not confined to parameter values used during fitting. Each modified representation is standardized and fitted separately, so these rows address different aspects of the observation and fitting protocol rather than estimating one common intervention.

\textbf{Observation and algebraic safeguards.} The preprocessing uses a standard-deviation floor of $10^{-6}$ and clips the standardized field to $[-10,10]$. Across all 23,760 saved candidate fields, the smallest standard deviation is 0.00527 and the largest absolute standardized value is 4.669, so neither safeguard is active. The calculation of $G_m$ uses a separate standard-deviation floor of $10^{-12}$, which is also inactive. The largest absolute fixed-field CHO residual in the logarithmic spectral rate is $6.66\times10^{-14}$, and the largest residual in the identity $(\mathcal O_m+\mathcal S_m)-[G_m[\phi_m]-G_0[\phi_0]]$ is $2.78\times10^{-17}$. These values check the implemented algebra; they do not establish continuum accuracy.

\begin{figure}[H]

\centering

\includegraphics[width=\textwidth]{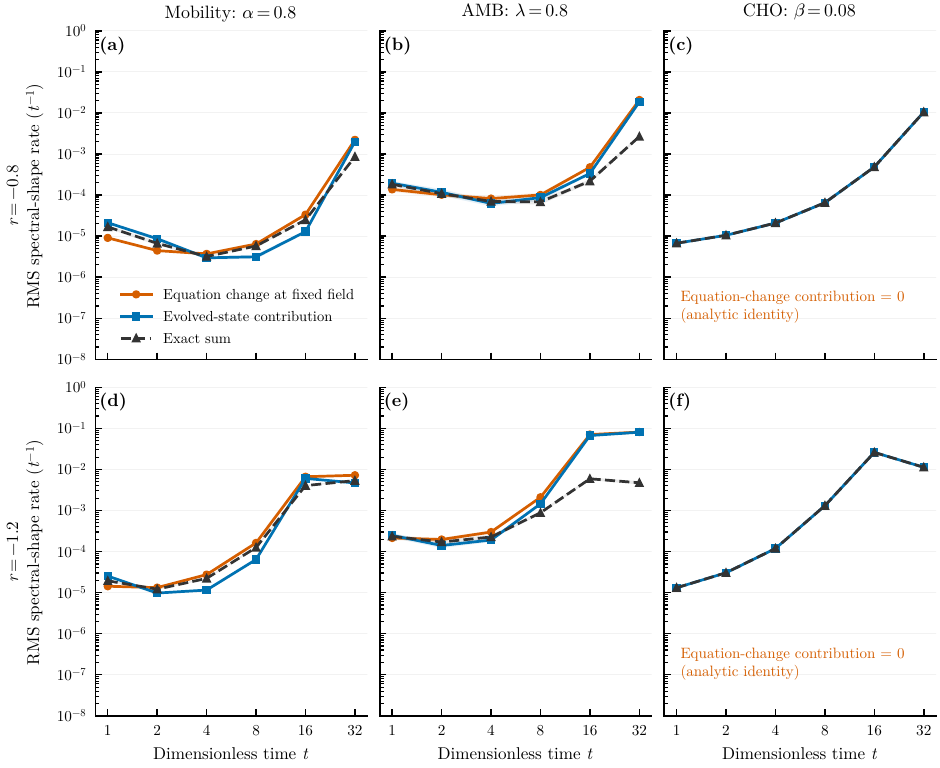}

\caption{\textbf{Direct equation and evolved-state contributions need not reinforce one another.} The strongest sampled Mobility ($\alpha=0.8$), AMB ($\lambda=0.8$), and CHO ($\beta=0.08$) trajectories are paired with Model B at identical initial fields. For each initial condition, curves show the root-mean-square magnitudes of $\mathcal{O}_m$, $\mathcal{S}_m$, and their exact sum across the 24 measured coordinates; shading gives pointwise 95\% intervals over 30 initial conditions. The CHO direct equation contribution is omitted from the logarithmic axis because it is analytically zero, so its evolved-state contribution and total coincide.}

\label{fig:budget-rms}

\end{figure}

\textbf{Time-step and nonlinear-grid checks.} The checks below compare numerical discrepancies with the physical difference between a candidate trajectory and its paired Model-B trajectory in the normalized spectral coordinates. For each initial condition, the candidate-minus-Model-B difference is a 24-component residual vector. We first average this vector over the four tested initial conditions and then take its root-mean-square across the 24 coordinates. This differs from Figure~\ref{fig:budget-rms}, where the root-mean-square is taken for each initial condition before averaging.

For CHO, two first-order integrators are compared: the exact-decay split used for the classification trajectories and an alternative update that places the restoring term in the denominator together with the implicit linear terms. Each integrator is extrapolated toward zero time step from $\Delta t=0.01$ and $0.005$ by taking twice the finer-step residual minus the coarser-step residual. The denominator in the CHO ratios is the norm of the average of the two extrapolated candidate-minus-Model-B residuals. Thus "time-step-limit difference" below means an estimate at the fixed $64\times64$ spatial discretization, not a joint space-time continuum limit.

For Mobility and AMB, the nonlinear products are also evaluated on Fourier grids enlarged by factors of 2 or 3 in each spatial direction: products are computed on $128\times128$ or $192\times192$ grids and then truncated back to the retained $64\times64$ modes. This padding reduces aliasing. In each padding ratio, the numerator is the difference between the two indicated candidate-minus-Model-B residuals, and the denominator is the norm of the residual computed with the finer padded grid relative to the unpadded Model-B reference at the same time step. A small physical candidate-minus-Model-B difference can therefore make a ratio large even when both absolute quantities are small. The checks cover four initial conditions, both quench depths, and the weakest and strongest parameter values through $T=16$; they do not cover the full classification population.

\begin{table}[H]
\centering
\small
\begin{tabularx}{\textwidth}{>{\raggedright\arraybackslash}X >{\centering\arraybackslash}p{0.20\textwidth} >{\centering\arraybackslash}p{0.25\textwidth}}
\toprule
\textbf{Numerical comparison} & \textbf{Maximum ratio at $T=16$} & \textbf{Maximum over sampled positive times through 16} \\
\midrule
CHO: difference between the two extrapolated integrator limits / average extrapolated candidate-minus-Model-B difference & 0.00135 & 0.726 \\
CHO: bias of the exact-decay split at $\Delta t=0.02$ / average extrapolated candidate-minus-Model-B difference & 0.0449 & 0.383 \\
Mobility: difference between 2x- and 3x-padded residuals / 3x-padded candidate-minus-Model-B difference & $1.46\times10^{-5}$ & $1.95\times10^{-4}$ \\
AMB: difference between unpadded and 2x-padded residuals / 2x-padded candidate-minus-Model-B difference & 0.098 & 0.701 \\
AMB: difference between 2x- and 3x-padded residuals / 3x-padded candidate-minus-Model-B difference & $4.14\times10^{-9}$ & $3.11\times10^{-8}$ \\
\bottomrule
\end{tabularx}
\caption{Numerical discrepancy relative to the paired candidate-minus-Model-B difference over the stated validation subset. The sampled positive times are 1, 2, 4, 8, and 16. The CHO calculations use time steps 0.02, 0.01, and 0.005. The padding rows include both tested time steps, 0.02 and 0.01. These quantities diagnose numerical sensitivity; they are not uncertainty intervals for classification accuracy.}
\label{tab:numerical}
\end{table}

The observed convergence order of the difference between the two CHO integrators, estimated from the 0.01-to-0.005 pair at $T=16$, ranges from 0.997 to 1.005 across the parameter and quench combinations. The corresponding estimate from the coarser pair ranges from 0.995 to 1.010. The close extrapolated limits at $T=16$ do not imply equally close agreement at every earlier time. The largest all-time ratio, 0.726, occurs at $t=1$ for $r=-0.8$ and $\beta=0.08$, where both the absolute integrator difference and the physical candidate-minus-Model-B difference are small. At $\Delta t=0.02$, the alternative denominator-based update has a worst $T=16$ bias-to-effect ratio of 1.29, compared with 0.0449 for the exact-decay split used to generate the present CHO trajectories.

For AMB, the near agreement between 2x and 3x padding shows numerical stability with respect to this refinement in the tested regime. It does not show that evaluating nonlinear products directly on the original $64\times64$ grid is already equivalent to the padded result: the unpadded-to-2x ratio remains 0.098 at $T=16$ and is larger at some earlier times. At a fixed 3x-padded AMB field, the root-mean-square difference between the rates of the measured 24-vector computed with 2x and 3x padding is at most $1.05\times10^{-9}$ at $T=16$ and $2.16\times10^{-9}$ over the sampled positive times, after division by the 3x-padded AMB-minus-Model-B rate difference at that same field. The norm of the difference between the $\Delta t=0.02$ and 0.01 candidate-minus-Model-B residual vectors is at most 0.0373 times the norm of the $\Delta t=0.01$ residual for Mobility and 0.0322 times it for AMB at $T=16$; these comparisons use 3x and 2x padding, respectively. None of these targeted tests establishes convergence of every classification boundary, especially at $T=32$.

\textbf{Selecting subsets of time-bin coordinates.} We next ask whether coordinates at which the three model families have different decomposition magnitudes are especially informative to the classifier. A coordinate is a pair $(t,j)$ consisting of a saved time and one of the 24 radial bins. The selection is repeated independently within each training split and observation time $T$, and $t=0$ is excluded because all paired trajectories share the same initial field.

For family $f$, define $B_f(t,j)$ as the square root of the training-trajectory mean of $\mathcal{O}_f(t,j)^2+\mathcal{S}_f(t,j)^2$, pooling both quenches and all four parameter values. Let $c_f$ be the root-mean-square of $B_f$ over all eligible $(t,j)$ coordinates and set $b_f=B_f/c_f$. The score

\begin{equation}
q(t,j)=\sqrt{\frac{1}{3}\sum_f\left(b_f(t,j)-\frac{1}{3}\sum_g b_g(t,j)\right)^2}.
\end{equation}

is large where these normalized magnitudes differ most among the three families. We retain either the highest-scoring 20\% or an equal number of the lowest-scoring coordinates, rounding the count to the nearest integer. In complementary tests, we delete the same high- or low-scoring coordinates from the full history. The retained-subset inputs exclude the common initial frame, whereas the deletion inputs keep it. Each resulting representation is standardized and fitted independently. Retaining and deleting answer different questions because information in a selected region can be duplicated elsewhere in the history.

\begin{table}[H]
\centering
\small
\begin{tabularx}{\textwidth}{>{\centering\arraybackslash}p{0.10\textwidth} >{\raggedright\arraybackslash}X >{\raggedright\arraybackslash}X}
\toprule
\textbf{Final observation time $T$} & \textbf{Retain high-score minus retain low-score (percentage points)} & \textbf{Delete high-score minus delete low-score (percentage points)} \\
\midrule
8 & 12.50 [8.89, 16.39] & -0.97 [-4.31, 2.22] \\
16 & 0.56 [-2.64, 3.75] & 1.25 [-1.53, 4.03] \\
32 & 8.19 [3.61, 12.78] & -2.36 [-5.56, 0.56] \\
\bottomrule
\end{tabularx}
\caption{Classification controls based on subsets of time-bin coordinates selected from the decomposition into direct equation and evolved-state contributions. The Mobility rate is evaluated with a constant-mobility baseline that matches the spectral Laplacian used by the time integrator, and the CHO rate uses the exact-decay update. Brackets give nominal pointwise 95\% percentile intervals from 10,000 resamples of the 30 initial-condition groups, with the selected coordinate subsets, preprocessing, regularization choices, and fitted classifiers held fixed. A positive retained-subset contrast means that the high-score subset is more useful than an equal-size low-score subset. A negative deletion contrast would mean that deleting the high-score subset is more harmful. All three deletion intervals contain zero.}
\label{tab:coordinates}
\end{table}

The selected masks are stable across nearly all fold--representation combinations. The small number of boundary changes does not affect the primary endpoint/history analysis. These subset results suggest where useful information occurs for this score and classifier; they do not identify coordinates that are necessary independently of the fitted model.

\textbf{Direction and order-of-change checks.} The cancellation index depends on both the relative direction and the relative magnitude of $\mathcal{O}_m$ and $\mathcal{S}_m$. We therefore compare the measured cancellation with two additional quantities, all computed in the 24-dimensional observed space at one saved time. Let $\mathcal{C}_{\mathrm{sym}}$ denote the cancellation index from the symmetric decomposition in Section~\ref{sec:diagnostics}. If two vectors with the same measured norms were orthogonal instead, their cancellation index would be

\begin{equation}
\mathcal{C}_{\perp}=1-\frac{\left(\|\mathcal{O}_m\|_2^2+\|\mathcal{S}_m\|_2^2\right)^{1/2}}{\|\mathcal{O}_m\|_2+\|\mathcal{S}_m\|_2}.
\end{equation}

For any nonnegative ratio of the two norms, $\mathcal{C}_{\perp}\leq1-1/\sqrt{2}\simeq0.293$, with equality only when the norms are equal. Cancellation above this bound therefore requires stronger opposition than orthogonal vectors with the same norms. Their direction is measured directly by the cosine

\begin{equation}
\rho_{\mathrm{op},state}=\frac{\langle\mathcal{O}_m,\mathcal{S}_m\rangle}{\|\mathcal{O}_m\|_2\|\mathcal{S}_m\|_2},
\end{equation}

where $-1$, 0, and 1 denote opposite, orthogonal, and aligned vectors. The cosine is undefined if either vector has zero norm.

The symmetric decomposition averages the two possible orders for changing the equation and the field. In the first order, the field is changed from $\phi_0$ to $\phi_m$ while the Model-B equation is held fixed, and the equation is then changed at $\phi_m$. The two exact contributions are

\begin{equation}
\mathcal{O}^{(A)}_m=G_m[\phi_m]-G_0[\phi_m],\qquad \mathcal{S}^{(A)}_m=G_0[\phi_m]-G_0[\phi_0].
\end{equation}

In the second order, the equation is first changed at the Model-B field $\phi_0$, and the field is then changed while the candidate equation is held fixed. The two contributions are

\begin{equation}
\mathcal{O}^{(B)}_m=G_m[\phi_0]-G_0[\phi_0],\qquad \mathcal{S}^{(B)}_m=G_m[\phi_m]-G_m[\phi_0].
\end{equation}

Each pair sums exactly to $G_m[\phi_m]-G_0[\phi_0]$. The symmetric terms are their averages: $\mathcal{O}_m=(\mathcal{O}^{(A)}_m+\mathcal{O}^{(B)}_m)/2$, with the corresponding identity for $\mathcal{S}_m$. We compute the same cancellation index for both ordered decompositions. Every quantity is formed separately for each realization before averaging over the 30 initial-condition groups. The reported range is the smaller and larger of the two mean cancellation values; it measures sensitivity to the order of comparison and is not a confidence interval.

\begin{table}[H]
\centering
\small
\begin{tabularx}{\textwidth}{>{\raggedright\arraybackslash}p{0.16\textwidth} >{\centering\arraybackslash}p{0.08\textwidth} >{\centering\arraybackslash}p{0.13\textwidth} >{\centering\arraybackslash}p{0.13\textwidth} >{\centering\arraybackslash}p{0.16\textwidth} >{\centering\arraybackslash}X}
\toprule
\textbf{Family} & \textbf{Time} & \textbf{$\mathcal{C}_{\mathrm{sym}}$} & \textbf{$\mathcal{C}_{\perp}$} & \textbf{$\rho_{\mathrm{op},state}$} & \textbf{$\mathcal{C}$ range across the two orders} \\
\midrule
Mobility & 8 & 0.446 & 0.227 & -0.703 & 0.442--0.450 \\
Mobility & 16 & 0.681 & 0.292 & -0.799 & 0.637--0.716 \\
Mobility & 32 & 0.547 & 0.275 & -0.664 & 0.505--0.585 \\
AMB & 8 & 0.752 & 0.281 & -0.938 & 0.350--0.880 \\
AMB & 16 & 0.956 & 0.293 & -0.998 & 0.797--0.979 \\
AMB & 32 & 0.970 & 0.293 & -0.998 & 0.874--0.986 \\
\bottomrule
\end{tabularx}
\caption{Direction and sensitivity to the order of comparison for the strongest Mobility and AMB perturbations at $r=-1.2$. Values are means over 30 paired initial-condition groups. CHO is omitted because $\mathcal{O}_m=0$ analytically, so the cosine between $\mathcal{O}_m$ and $\mathcal{S}_m$ is undefined. Its zero cancellation index is the limit in which only the evolved-state contribution remains; it is not a no-cancellation reference comparable to the other families.}
\label{tab:decomposition-sensitivity}
\end{table}

At late times, the AMB direct equation and evolved-state contributions remain more strongly opposed than the orthogonal reference for both orders of comparison. Mobility also shows greater opposition than its orthogonal reference, but the effect is weaker and less persistent.

\section{Dataset and classifier support}\label{app:classifier-support}

Table~\ref{tab:population} summarizes the classification population and the independent units used for splitting and resampling. The trajectory count includes only candidates with positive perturbation parameters. Model-B trajectories are retained as paired physical references and are not assigned a classification label.

\begin{table}[H]
\centering
\small
\begin{tabularx}{\textwidth}{>{\raggedright\arraybackslash}X >{\raggedright\arraybackslash}p{0.34\textwidth}}
\toprule
\textbf{Quantity} & \textbf{Value} \\
\midrule
Independent initial-condition groups & 30 \\
Quench depths & 2 ($r=-0.8,-1.2$) \\
Candidate models & 3 \\
Positive parameter values per model & 4 \\
Candidate trajectories per initial-condition group & 24 \\
Candidate trajectories with positive parameter values & 720 \\
Paired Model-B reference trajectories & 60 \\
Saved frames per trajectory & 33, including $t=0$ \\
Grouped train/validation/test splits & 5 \\
Train/validation/test groups per split & 18/6/6 \\
Bootstrap units & 30 initial-condition groups \\
Bootstrap resamples & 10,000 \\
CHO trajectories with positive $\beta$ using the exact-decay substep & 240 \\
\bottomrule
\end{tabularx}
\caption{\textbf{Analysis population and independent statistical units.} All model families, parameter values, and quenches descended from one initial field remain in the same training, validation, or test partition. The 720 classification trajectories equal $3\times4\times2\times30$. The paired Model-B trajectories are used only for physical comparisons.}
\label{tab:population}
\end{table}

Table~\ref{tab:all-horizons} gives the complete endpoint/history comparison for all five final observation times. At each time, the accuracies pool held-out predictions from the same five grouped train/validation/test splits.

\begin{table}[H]
\centering
\small
\begin{tabularx}{\textwidth}{>{\centering\arraybackslash}p{0.11\textwidth} >{\centering\arraybackslash}p{0.19\textwidth} >{\centering\arraybackslash}p{0.19\textwidth} >{\centering\arraybackslash}X}
\toprule
\textbf{Final observation time $T$} & \textbf{Endpoint accuracy (\%)} & \textbf{History accuracy (\%)} & \textbf{History minus endpoint (percentage points)} \\
\midrule
2 & 58.89 & 56.67 & -2.22 [-4.86, 0.42] \\
4 & 54.86 & 57.64 & 2.78 [-1.11, 6.81] \\
8 & 54.44 & 58.61 & 4.17 [0.69, 7.64] \\
16 & 48.06 & 54.44 & 6.39 [3.75, 8.89] \\
32 & 56.11 & 61.81 & 5.69 [1.53, 9.72] \\
\bottomrule
\end{tabularx}
\caption{\textbf{Complete endpoint/history comparison.} Brackets give pointwise 95\% percentile intervals from 10,000 resamples of the 30 initial-condition groups, with the fitted preprocessing and classifiers held fixed. The comparisons among final observation times in Section 3.5 use the stated Holm adjustment; the intervals shown here are descriptive and are not simultaneous confidence intervals.}
\label{tab:all-horizons}
\end{table}

Fixed-prediction parameter subsets show why the main result should not be read as uniform over perturbation strength. At $T=16$, evaluation on only the weakest positive parameter level in each family gives a history-minus-endpoint difference of -2.22 points [-6.67, 2.78] across 180 trajectories, whereas excluding those weakest levels gives 9.26 points [5.74, 12.59] across the remaining 540 trajectories. These comparisons reuse the preprocessing and classifiers fitted to the full positive-parameter population; the stricter refitted leave-one-level-out result is reported in Appendix~\ref{app:checks} and Figure~\ref{fig:controls}.

\begin{figure}[H]

\centering

\includegraphics[width=\textwidth]{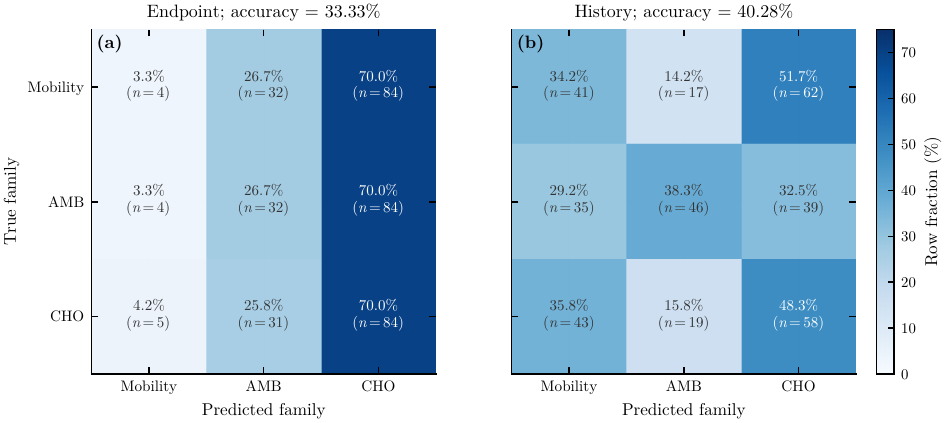}

\caption{\textbf{Shallow-quench confusion matrices at $T=16$.} Rows are true families and columns are predicted families for the held-out $r=-0.8$ population, with 120 trajectories per family and 360 trajectories in total. Each cell gives its fraction of the corresponding row and its count. The classifier using only the last 24-vector has accuracy 33.33\%, whereas the classifier using all saved 24-vectors through $T=16$ has accuracy 40.28\%. The matrices expose asymmetric errors of these fixed preprocessing procedures and logistic classifiers. They neither establish that the three endpoint distributions are identical nor imply unique recovery of a governing equation.}

\label{fig:confusion-shallow}

\end{figure}

The endpoint accuracy at $T=16$ for the shallow quench is exactly one third, but that aggregate value does not show which models are confused. Figure~\ref{fig:confusion-shallow} resolves those class-specific errors for the two input representations.

\section{Additional decomposition results}\label{app:mechanism-support}

\begin{figure}[H]

\centering

\includegraphics[width=\textwidth]{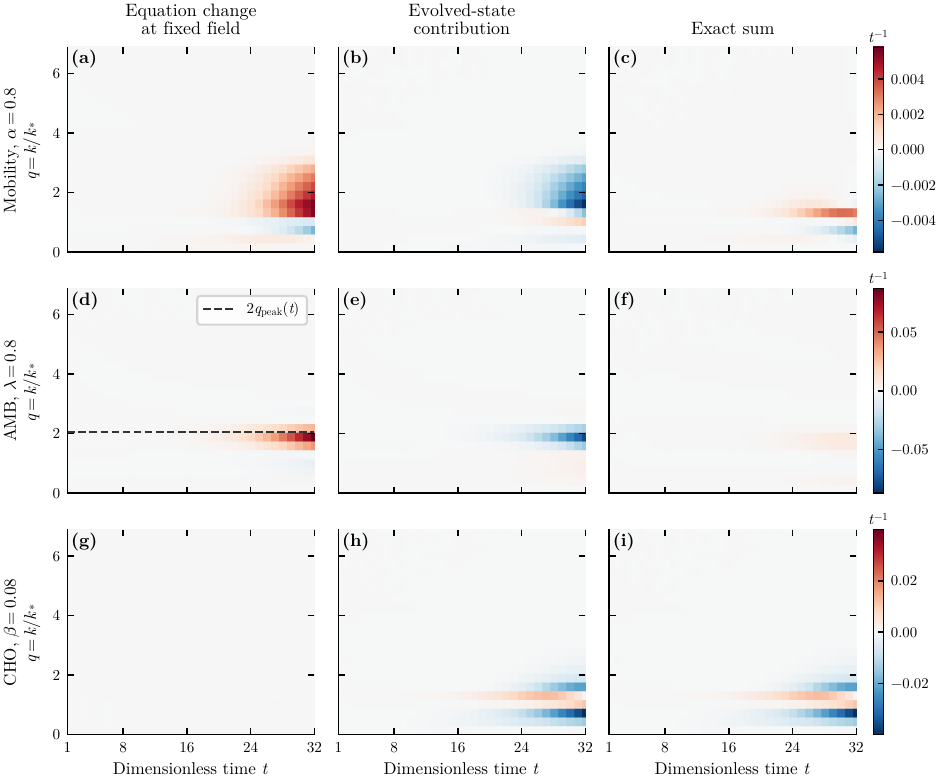}

\caption{\textbf{Direct equation and evolved-state contributions for the shallow quench.} The construction matches Figure~\ref{fig:signed-budget} but uses $r=-0.8$ and the strongest sampled Mobility ($\alpha=0.8$), AMB ($\lambda=0.8$), and CHO ($\beta=0.08$) perturbations. Each map is the signed mean over 30 paired initial-condition groups at the saved dimensionless times $t=1,\ldots,32$. The vertical coordinate is $q=k/k_*$, where $k_*=\sqrt{|r|/(2\kappa)}$ is the wave number with the largest linear Model-B growth rate. The three columns show $\mathcal{O}_m$, $\mathcal{S}_m$, and their exact sum, all in units of $t^{-1}$. Each family row uses its own symmetric color scale. In the AMB row, the dashed curve is $2q_{\mathrm{peak}}(t)$, where $q_{\mathrm{peak}}=k_{\mathrm{peak}}/k_*$ and $k_{\mathrm{peak}}$ is the average of the candidate and reference spectral peak-bin centers. These maps describe instantaneous physical rates and are not classifier inputs.}

\label{fig:signed-budget-shallow}

\end{figure}

The signed maps in the main text use the deeper quench, $r=-1.2$. Figure~\ref{fig:signed-budget-shallow} repeats the same calculation for the shallower quench, $r=-0.8$, so that the two regimes can be compared directly.

The AMB guide follows from the Fourier-space form of its quadratic-gradient term. Let $u=|\nabla\phi|^2$, let $f_{\lambda}[\phi]=\lambda\nabla^2u$, and define $\widehat{\phi}^{\,c}_{\mathbf{k}}=\int_{\Omega}\phi(\mathbf r)e^{-i\mathbf k\cdot\mathbf r}\,d\mathbf r$. Then

\begin{equation}
\begin{aligned}
\widehat{u}^{\,c}_{\mathbf{k}}
&=-\frac{1}{|\Omega|}\sum_{\mathbf{q}}
\mathbf{q}\cdot(\mathbf{k}-\mathbf{q})\\
&\quad\times
\widehat{\phi}^{\,c}_{\mathbf{q}}
\widehat{\phi}^{\,c}_{\mathbf{k}-\mathbf{q}},
\end{aligned}
\end{equation}

\begin{equation}
\widehat{f}_{\lambda,\mathbf{k}}^{\,c}
=-\lambda k^2\widehat{u}_{\mathbf{k}}^{\,c}.
\end{equation}

The active term therefore combines pairs of Fourier modes. When most power lies near $k_{\mathrm{peak}}$, sums of dominant wave vectors can reach about $2k_{\mathrm{peak}}$. Angular averaging, finite radial bins, and the discrete grid broaden this relation, so it predicts a band rather than an exact harmonic.

The band in the AMB direct equation contribution at the deeper quench can be located at the finite resolution of the 24 radial bins. Let $q_{\mathrm{op}}=k_{\mathrm{op}}/k_*$ denote the bin center at which the absolute ensemble-mean direct equation contribution is largest, and let $q_{\mathrm{peak}}=k_{\mathrm{peak}}/k_*$ denote the average of the candidate and Model-B spectral peak locations defined above. Table~\ref{tab:amb-ridge} compares $q_{\mathrm{op}}$ with $2q_{\mathrm{peak}}$ for the strongest AMB perturbation.

\begin{table}[H]
\centering
\small
\begin{tabularx}{\textwidth}{>{\centering\arraybackslash}p{0.07\textwidth} >{\centering\arraybackslash}p{0.16\textwidth} >{\centering\arraybackslash}p{0.18\textwidth} >{\centering\arraybackslash}p{0.18\textwidth} >{\centering\arraybackslash}X}
\toprule
\textbf{Time} & \textbf{Direct-equation peak $q_{\mathrm{op}}$} & \textbf{Spectrum peak $q_{\mathrm{peak}}$} & \textbf{$q_{\mathrm{op}}/q_{\mathrm{peak}}$} & \textbf{Relative difference from $2q_{\mathrm{peak}}$ (\%)} \\
\midrule
8 & 2.031 & 1.075 & 1.889 & 5.56 \\
16 & 1.792 & 0.836 & 2.143 & 7.14 \\
32 & 1.553 & 0.836 & 1.857 & 7.14 \\
\bottomrule
\end{tabularx}
\caption{\textbf{Location of the AMB direct-equation band for $r=-1.2$ and $\lambda=0.8$.} The direct equation contribution and the contemporaneous candidate and Model-B spectra are averaged separately over the 30 paired initial-condition groups before their peak bins are located. The last column is $100\lvert q_{\mathrm{op}}-2q_{\mathrm{peak}}\rvert/(2q_{\mathrm{peak}})$. At the reported times, the peak of the direct equation contribution lies within approximately one radial bin of $2q_{\mathrm{peak}}$. This is consistent with quadratic mode coupling at the available 24-bin resolution; it is not an exact harmonic identity.}
\label{tab:amb-ridge}
\end{table}

\end{document}